# Students' understanding of direct current resistive electrical circuits

by
Paula Vetter Engelhardt and Robert J. Beichner
Department of Physics, North Carolina State University, Raleigh, North Carolina 27695

## Abstract

Both high school and university students' reasoning regarding direct current resistive electric circuits often differ from the accepted explanations. At present, there are no standard diagnostic tests on electric circuits. Two versions of a diagnostic instrument were developed, each consisting of 29 questions. The information provided by this test can provide instructors with a way of evaluating the progress and conceptual difficulties of their students. The analysis indicates that students, especially females, tend to hold multiple misconceptions, even after instruction. During interviews, the idea that the battery is a constant source of current was used most often in answering the questions. Students tended to focus on the current in solving problems and to confuse terms, often assigning the properties of current to voltage and/or resistance.

## I. Introduction

In recent years, physics educators have begun to look more closely at what their students understand about physics concepts. Students' patterns of response to questions about circuit phenomena often are in conflict with those accepted by the physics community. The term "misconception" will be used to refer to students' incorrect pattern of response. This pattern could be part of a coherent naive theory of some physical phenomena or a more fragmented and primitive response produced on the spot as a result of the questions posed.

Widespread use of test instruments such as the Force Concept Inventory (FCI)[1] and the Test of Understanding Graphs in Kinematics (TUG-K)[2] has brought a new way of evaluating students' conceptual understanding. However, more instruments need to be developed in a variety of areas to allow instructors to better evaluate their students' understanding of physics concepts and to evaluate new teaching endeavours for their feasibility. The Determining and Interpreting Resistive Electric Circuit Concepts Test (DIRECT) was developed to evaluate students' understanding of a variety of direct current (DC) resistive electric circuits concepts. DIRECT has been designed for use with high school and college/university students. Common misconceptions were incorporated into the distracters of the test items.

We will discuss the development of DIRECT versions 1.0 and 1.1 and will examine their feasibility for assessing students' conceptual understanding and potential use in evaluating curricula. We will answer the following research questions: (1) Can a multiple-choice test be developed that is reliable, valid, and uncovers students' misconceptions? (2) Are there significant differences between various groups of students taking DIRECT? In particular, are there noticeable differences between course level (high school versus university), gender, and instructional methods? (3) What misconceptions can the test detect?

The body of knowledge regarding students' understanding of DC resistive electric circuits is quite extensive.[3] Students' typical response patterns indicate that they make two assumptions regarding DC resistive electrical circuits: current is consumed,[4] and the battery is a source of constant current.[5] In addition, students interchangeably use terms associated with circuits, often assigning the properties of current either to voltage, resistance, energy, or power.[6]

Physicists use schematic diagrams to represent circuit elements and examine their behavior. Students' recognition of what these diagrams represent is an important aspect of their understanding of circuits. Research reveals that students view these diagrams as a system of pipes within which flows a fluid that they refer to as electricity.[7] Students have difficulty identifying series and parallel connections in diagrams.[8] Students do not understand and do not correctly apply the concept of a complete circuit.[9] Gott[10] has reported that more than 90% of students age 15 recognized the need for a complete circuit. However, he found a small but significant group of students who would include a short circuit (such as a shorted battery) as an acceptable complete circuit.

In analyzing circuits, students view it in a piece-meal fashion in contrast to a global view. There is some evidence[11] to indicate that students change their reasoning patterns to suit the question at hand. Thus, they do not appear to use a single, consistent model to analyze circuit phenomena. Instead, students use one of three ways of reasoning: sequential, local, or superposition. Sequential reasoning results in a “before and after” examination of the circuit. Students using sequential reasoning believe that current travels around the circuit and is influenced by each element as it is encountered, and a change made at a particular point does not affect the current until it reaches that point.[12] Thus, for the circuit shown in Fig. 1, closing the switch will not affect bulb A because the current has already passed that point. Von Rhöneck and Grob differentiate local from sequential reasoning in the following way: “local reasoning means that the current divides into two equal parts at every junction regardless of what is happening elsewhere.”[13] Given the circuit shown in Fig. 2, students would say that the current in branch 1 was equal to that in branch 2. Students using superposition reasoning would conclude that if one battery makes a bulb shine with a certain brightness, then two batteries would make the bulb shine twice as bright, regardless of the configuration.[14]

When confronted with a qualitative problem, students show reluctance when asked to reason qualitatively and resort to technical or quantitative approaches.[15] This reluctance is said to be due to a lack of experience solving qualitative problems.[16] Additionally, students have been shown to have difficulty mastering reasoning with ratios.[17]

Tests on DC resistive electric circuits do exist,[18] but they have mostly been developed as either a research tool or curriculum assessment instrument, not as a general assessment tool. Thus, there are limitations with many of these tests that prevent them from being used for this purpose. Those that have been developed as a research tool often have restricted content, looking at a single concept such as resistance.[19] Those that do cover more topics generally have a single item for each objective,[20] which does not allow for comparisons between questions nor provide additional statistical evidence of comprehension. (Was it the question or the concept that students didn't understand?) Statistical evidence pertaining to the reliability and validity of the tests has not been well documented. Many of the assessment tests were developed mainly to evaluate and to revise the curriculum materials with which they were associated. Although some of these tests reveal and quantify students' conceptual understanding,[21] they usually were not intended to be used in a wider format. Many of these tests have been administered to small groups of students with similar abilities or only to the groups under investigation. Small sample sizes can increase the sampling error.

Thus, a test that could be used as both a research tool in assessing new curriculum materials or teaching strategies as well as evaluating students' conceptual views that has sound statistical evidence of its reliability and validity is needed for DC resistive circuits.

## II. Development of DIRECT versions 1.0 and 1.1

As a first step in developing DIRECT, a set of instructional objectives was constructed after an extensive examination of high school and university textbooks and laboratory manuals plus informal discussions with instructors using those materials. The objectives were presented to a panel of independent experts to ensure that no fundamental concepts were overlooked. The final objectives are shown in Table I.

One typical comment that the panel made regarding the objectives was the omission of the use of meters in terms of their placement in circuits and their use as a measurement device to determine the behavior of the circuit. Although an important part of laboratory work, meters serve as an application of electric circuits concepts in contrast to a distinct concept of their own. Research has shown that students fail to treat meters as circuit elements and to recognise the implications for their construction and external connections.[22] Psillos, Koumaras, and Valassiades[23] found that a group of 14-15 year old Greek students believed that an ammeter would consume current so that it functioned in the same manner as a light bulb. The students did not understand that a good ammeter simply allows current to flow through it and has a negligible effect on the circuit. Thus, if such devices were included in the test, it would be difficult to determine if students were having difficulties with circuit concepts like current, or if they were having difficulties with the use and function of the meters.

The test was developed first in an open-ended format so that distracters for the multiple-choice version could be constructed. Efforts were made to write several items per objective. For example, three questions using a different mode of representation were written for objective 5. The three modes were verbal to schematic, realistic to schematic, and schematic to realistic. Some test items were adapted from the *Physics by Inquiry*[24] materials and *College Physics*[25] by Serway and Faughn. Members of the independent panel of experts suggested some items; however, most of the items were original.

In general, the questions were not aligned with any particular instructional approach so that the results would be the applicable to the largest possible audience. Questions written for objective 9, microscopic aspects of circuits, were the only exception and were closely aligned with the approach proposed by Chabay and Sherwood in their text, *Electric and Magnetic Interactions.*[26] They were included to evaluate how well students understand the microscopic aspects of circuits as this connection has only recently begun to be explored in some of the newer textbooks. As Cohen, Eylon and Ganiel[27] have noted, this lack of a causal relation may be the cause of some of the problems students have with electric circuits.

Large sample sizes were desired to reduce the magnitude of sampling error.[28] Thus, test sites were solicited via a message placed on a listserv for physics education researchers and educators (PHYS-LRNR) requesting test sites for the multiple-choice versions of the instrument and via contacts made during the 1993 Physics Courseware Evaluation Project's (PCEP) Summer Teachers' Institute held at North Carolina State University.

The multiple-choice version 1.0 of DIRECT (given in Appendix A) was administered to 1135 students from high schools ($N$ = 454) and universities ($N$ = 681) across the United States. The 29-item test took approximately half an hour to complete. The statistical analysis of the test is presented in Table II along with information about the statistics and their ideal values. Figure 3 shows the distribution of scores for the total sample, which is positively skewed, indicating a difficult test. Table III shows the percentage of students selecting each answer choice for each question as well as the point bi-serial correlation, discrimination, and difficulty of each question.

DIRECT version 1.1[29] was developed after an analysis of the results as well as individual follow-up interviews indicated that DIRECT version 1.0 needed to be revised to improve its reliability as well as to clarify questions that were confusing to students. There were two main revisions. The first was to increase the number of answer choices to 5 for all questions. In so doing, some questions became more quantitative in nature, asking by how much the brightness changed in contrast to asking if it increased / decreased or remained the same. The second was to redraw the circuit diagrams containing a light bulb in a socket using only the battery, bulb, and wires as the interviews indicated that students were confused about this representation.

DIRECT version 1.1 was administered to 692 students from high schools ($N$ = 251) and universities ($N$ = 441) in Canada (one high school and one university test site), Germany (one high school test site), and the United States. Version 1.1 consisted of 29 items, each with 5 answer choices, and took approximately half an hour to complete. The statistical analysis of the test is presented in Table II. Figure 3 shows the distribution of scores for the total sample, which also are positively skewed, indicating a difficult test. Table IV shows the results for version 1.1 in a similar manner to that of Table III.

## III. General findings

We will next discuss the discrimination ability (how well a particular question differentiated between students scoring well and students scoring poorly on the test) and how well students performed on the overall objectives listed in Table I for each version of the test.

## IIIA. Discrimination

Discrimination is a measure of the ability of a question to differentiate between students who scored well overall on the test from those who did not. Examining the data from version 1.0 revealed that question 26 was the most discriminating. To answer this question correctly, students could not reason sequentially, believe that the battery was a constant source of current, or think that current was consumed.

For the overall sample (combined university and high school) and for the university sample, questions 20 and 28 were the least discriminating; even students who scored well overall on the test had difficulties with these questions. Question 20 deals with what causes a current in a bulb filament. Students confused cause and effect, choosing the option that the current caused the field. Question 28 deals with the concept of the battery as a source of constant potential difference. Many students reasoned that because the current in a part of the circuit is zero, the voltage also is zero. For the high school sample, question 18 was the least discriminating. This question shows four circuits containing a battery, some connecting wires, and a light bulb in a socket. Students were able to identify complete circuits, but were unable to eliminate those that contained shorts.

The discrimination indices for version 1.1 revealed that for the overall and the university sample, question 14 was the most discriminating. Students who answered correctly had to understand how to calculate the equivalent resistance for resistors in a series/parallel combination and to compare the equivalent resistance to that of two resistors in series. Question 27 was the most discriminating for the high school sample, and explores students' understanding of objectives 1-3 in Table I. For all samples (overall), question 11 proved the least discriminating, and examines the students' understanding of the microscopic aspects of current.

**IIIB. Performance on the objectives**

Table I shows how well students performed on each of the instructional objectives for both versions 1.0 and 1.1. An examination of the distracters of both versions showed that 17% of the students could not identify a short in a circuit and/or determine what effect the short had on the circuit, 10% did not know where the contacts are on a light bulb, 6% had trouble identifying a complete circuit, and 28% exhibited current/voltage confusion.

On both versions of DIRECT, students were able to translate from a realistic representation of a circuit to the schematic, but had more difficulty in identifying the correct schematic from a written description of the circuit or in identifying the correct realistic representation of a circuit from a schematic. In general, students could identify a complete circuit. The difficulty arose when students were asked to determine whether the circuit worked or not, often including circuits that contained shorts.

## IV. Can a multiple-choice test be developed that is reliable and valid and in addition uncover students' misconceptions?

For a test to be useful, it must be both reliable and valid. Reliability is an indication of how precisely we made the measurement or how consistently the test measures what it measures. The Kuder-Richardson formula 20 (KR-20) was used to evaluate the reliability of both versions of DIRECT. The KR-20 should be at or above 0.70 for group measurements. Although this was the case for both versions (see Table II), the somewhat low values could be the result of the low discrimination and high difficulty indices. The low average discrimination values may indicate that the test is indeed uncovering students' misconceptions.

The other important and vital characteristic of any test is its validity – the ability of the test to measure what it is intended to measure or the test's accuracy. Validity is not a quality that can be established in a single measurement, but is accumulated via several measurements. Content validity (Does the test cover the appropriate material?) was established by presenting the test and objectives to an independent panel of experts to insure that the domain was adequately covered. The panel took the test and matched test items with objectives. This process yielded a percentage agreement for the answer key as well as for the objectives. Both open-ended questions (during the early development stages) and multiple-choice questions were directed to the panel. In cases where agreement on the objectives was low, the questions were rewritten. Although each question was written to address a particular objective, the test involves items that require the test taker to utilize additional information not specifically asked by the question and hence some questions by necessity addressed more than one objective.

The construct validity (Does the test measure electric circuits' concepts like current, voltage, etc?) of DIRECT was evaluated through a factor analysis, which will only be discussed briefly here, and interviews. A factor analysis analyzes the interrelationships within the data and can be used to select groups of items that appear to measure the same idea or factor. The factor analysis performed for both versions used the Little Jiffy method which revealed 8 factors associated with version 1.0 and 11 factors associated with version 1.1.[30] The interviews served (1) to determine if the questions were being understood in ways that were not intended and to better understand students' choices and (2) to provide evidence of the test's construct validity by the replication of results from previous studies.

Individual follow-up interviews using a subset of 10 questions from version 1.0 with 17 university and 11 high school students were conducted as part of the construct validity check. These interviews provided information on whether the questions were being understood in ways contrary to what was intended. Each interview lasted approximately 30 to 40 minutes and was audio taped and later transcribed. Any notes that students made during the interview were collected. The interview was semi-structured and made use of a think-aloud procedure, which required students to verbalize aloud their thoughts as they emerged. The interview was divided into three parts: identification of symbols used on the test, definition of terms used on the test, and answering the test items, providing reasoning behind their choice and their confidence on their answer. The student's answers to the multiple-choice test were available to the interviewer during the interview. If students changed their answers from the multiple-choice test, they were asked to recall what their reasoning was when they answered the test originally. To ensure a uniform coding of the interview transcripts, another researcher was asked to code the transcripts. The reliability of the coding between the two researchers was established with 15% of the sample at each level (high school and university) with a percentage agreement of 88%.

The interviews showed that nearly all of the students understood the symbols used on the test with the exception of the light bulb in a socket; two-thirds knew that a light bulb had two connections; and one-third believed that there was only one connection which was located at the bottom of the bulb.

The interviews were able to replicate results of previous studies. For example, some students who chose option E on question 3 reasoned via battery superposition, replicating the results of Sebastià.[11] The following is an example of a student using the battery as a superposition idea for question 3. The student in the excerpt was enrolled in a traditional, calculus-based course.

> "I think I would put E because the batteries are providing the energy so since they both have two two [sic] batteries. I didn't think that it would matter whether they were in parallel or series because they're gonna add a certain amount of voltage and when the parallel batteries link up it's gonna be equivalent to whatever voltage is added when they are in series and then the light bulbs since they are just two in series, that's the same for all three pictures."

In reviewing the results obtained from the follow-up interviews with version 1.0, there initially appeared to be no pattern to the students' reasoning on the interviewed questions. However, examining which misconception was used most often on each question and comparing them with the global objectives (see Table V) for each question did yield a pattern. Table V shows the four main divisions or global objectives: physical aspects of the circuit, energy, current, and potential difference (voltage), and the misconceptions that were cued for the interview questions posed. For the global objective of voltage, the dominant misconceptions for these questions were battery as a constant current source, term confusion *I with V*, local reasoning, and battery superposition. These misconceptions relate to students' understanding of the properties of the battery and what it supplies to the circuit. Similarly, for the global objective of physical aspects of the circuit, typical misconceptions were topology, contacts, and term confusion *I/R*. These misconceptions related to the physical features of the circuit.

The topological errors indicated that students looked at the surface features of the circuit. The contact error indicated that students were missing some knowledge of where the contacts are located on a light bulb. Term confusion *I with R* errors indicated that students did not understand that a resistor (including light bulbs) has an inherent resistance based on its shape and the material from which it is made. One could categorize errors associated with the physical aspects of the circuits as students not having the declarative knowledge needed to understand the physical nature of the circuit diagram and its associated elements. Thus, although different questions cued the use of different misconceptions, the students tended to use misconceptions associated with the global objective of the question.

To summarize, there is evidence that both versions of DIRECT are reliable and valid. Both versions appear to be able to illicit students' conceptual understanding of DC resistive electric circuits concepts.

### V. Are there significant differences between level (High School versus University), gender, and instructional methods?

To answer this question, a series of *t*-tests and ANOVA were used to determine if there were significant differences between various groups of students who had taken DIRECT versions 1.0 and 1.1. Groups were considered significantly different if the level of significance or *p*-value was at or below 0.05, which gives a 95% level of confidence that the difference is real. All *t*-tests assumed a one-tail test of significance so that the superiority of one group over the other could be determined. Students' raw scores were used in these calculations, so that a score of 29 is equivalent to 100%.

### VA. Level (High school compared to university)

For version 1.0, there were significant differences in the averages for the university ($M$ = 15) and high school groups ($M$ = 12), $t$ (1008) = 11, $p < 3.8 \times 10^{-28}$, with university students outperforming high school students. There were no significant differences between calculus-based ($M$ = 16) and algebra-based ($M$ = 15) university students, $t$ (191) = -1.6, $p < 0.06$. No significant differences were found between the Advanced Placement or honors high school students ($M$ = 12) and those high school students taking a regular physics class ($M$ = 13), $t$ (342) = -0.89, $p < 0.19$. Similar results were obtained for version 1.1. The analysis of interview results found no significant differences in the number of misconceptions used by university ($M$ = 8) and high school students ($M$ = 9), $t$ (23) = -0.73, $p < 0.24$. However, university students were significantly ($p < 0.006$) more confident in their interview answers than were the high school students.

### VB. Gender

For version 1.0, significant differences were found in the averages for males and females with males outperforming females at all levels (see Table VI).

Interview results indicated significant differences between the number of misconceptions used by males ($M$ = 6) and females ($M$ = 11), $t(25) = 3.9$, $p < 0.0003$, with females using more than males. A similar finding was found for university males ($M$ = 6) and females ($M$ = 11), $t$ (11) = 3.6, $p < 0.002$. However, there were no significant differences found between high school males ($M$ = 6) and females ($M$ = 10), $t$ (4) = 1.4, p < 0.12. Males were more confident in their interview responses than were females ($p < 0.0006$).

**VC. Instructional method**

To evaluate the feasibility of using DIRECT to evaluate curricular materials and to assess new teaching methods, several subgroups who took DIRECT 1.0 and 1.1 were chosen for further examination. Part of the DIRECT 1.0 university sample contained a small group of calculus-based students who used a Chabay and Sherwood text[26], which discusses the microscopic aspects of circuit phenomena. We found that there were significant differences between students using the Chabay and Sherwood text ($M$ = 18) and students using more traditional textbooks ($M$ = 15), $t$ (76) = -3.8, $p < 0.0001$, as well as the university group as a whole (algebra and calculus-based combined) ($M$ = 15), $t$ (44) = -4.2, p $< 6.1 \times 10^{-5}$. Those students using the Chabay and Sherwood textbook outperformed both groups.

There was a small group of students who used the *Physics by Inquiry* materials, which uses an inquiry approach to instruction with many hands-on activities. This small group of students took DIRECT version 1.1. An analysis of variance (ANOVA) was performed which allows one to compare the means of more than two groups. Our results showed that there were significant differences between the students using the *Physics by Inquiry* materials ($M$ =15), calculus-based students ($M$ =13), and algebra-based students ($M$ =12), $F$ (2, 438) = 4.13, $p < 0.017$. Those students using *Physics by Inquiry* outperformed both groups.

This examination of various subgroups that used new curricular materials showed that there were statistically significant differences between their scores and students who were taking more traditional courses. These results are only preliminary and were performed to evaluate if DIRECT could be used in this way. More rigorously designed studies would need to be developed to further evaluate the apparent differences between these subgroups and other students. DIRECT appears to be able to assess differences between groups of students using differing instructional methods and materials.

## VI. What misconceptions can the test detect?

We now discuss the difficulties and misconceptions that DIRECT can detect. The interview results showed a variety of difficulties students experienced with a subset of questions from DIRECT 1.0 as shown in Table V.

A comparison of students' definitions of terms used on DIRECT and the student misconceptions indicates that the main source of the difficulty is with term confusion, generally associated with current. Students assign the properties of energy to current, and then assign these properties to voltage and resistance. Specifically, both voltage and resistance can only occur in the presence of a current.

Students do not have a clear understanding of the underlying mechanisms of electric circuits. This misunderstanding is most likely the result of a weak connection between electrostatics and electrokinetics phenomena, as this connection is only now beginning to be addressed in some of the newer textbooks.

Students were able to translate easily from a realistic representation of a circuit to the corresponding schematic diagram. Students had difficulty making the reverse translation. However, this result may be more indicative of their difficulty identifying shorts within circuits or of deficiencies in their knowledge regarding the contacts for light bulbs.

One aspect of DIRECT that sets it apart from other tests that have been developed is the use of batteries connected in series or parallel. This inclusion allows one to investigate how students interpret voltage and current in circuits containing these elements. Results from version 1.0 indicated that students had difficulty predicting the resulting voltage and current. Interviews indicated that some of the students were using superposition reasoning, while others were using a combination of battery as a constant current source and local reasoning. Hand-written notes made by the students during the interviews indicated that some students may have been trying to apply rules for equivalent resistors or capacitors to the battery arrangements. Version 1.1 explored further distinctions between two batteries in series and two batteries in parallel through questions 3 (in its original form) and question 7 (see Fig. 4). Results from these questions indicated the following:

1) Students who believe that two batteries in parallel provide more energy (27%) also believe that they provide more voltage (21%) (Pearson $r = 0.37$).

2) Students who believe that two batteries in series provide more energy (46%) also believe that they provide more voltage (51%) (Pearson $r = 0.45$).

3) Students who believe that two batteries in series and two batteries in parallel provide the same energy (17%) also believe that they provide same voltage (22%) (Pearson $r = 0.41$).

Those questions containing multiple batteries were items questioned by the independent panel of experts. They were concerned that this use might diminish the results of the test because multiple batteries are not typically taught. However, the ideas necessary to analyze these circuits are presented in most courses. The ideas are that the potential difference in two parallel branches remains the same while the currents in the parallel branches add to equal the total current available, and the potential difference across each element in series adds to equal the total input from the battery while the current remains the same. These ideas are used in a number of the problems and were acknowledged by the panel of experts as important to include on the test. Thus, if students truly understand these concepts, they should be able to apply them to novel situations.

## VII. Conclusions and Implications

Both versions of DIRECT appear to be reliable and valid. Results indicate that either version could be useful in evaluating curriculum or instructional methods as well as providing insight into students' conceptual understanding of DC circuit phenomena.

Interview results indicated that students use the idea that the battery is a constant current source most often in solving the interview problems. Students were found to use different misconceptions depending on the problem presented. Thus, different questions cued different misconceptions. Although students tended to use different misconceptions for each question presented, they did tend to use misconceptions related to the global objective of the question.

There are differences associated with gender in terms of performance, number of misconceptions used, and confidence and with course level with regard to performance and confidence. Generally, males outperformed females and had more confidence in their responses than did females. Females tended to use more misconceptions. Performance differences were found on both versions of DIRECT with university students outperforming high school students. University students also had more confidence in their answer selections.

In revising DIRECT 1.0, the number of answer choices was increased to five for all questions. In so doing, some questions became less qualitative and more quantitative. Instead of asking does the brightness increase, decrease, or stay the same, the questions asked by how much the brightness changed (1/4, 1/2, 2, 4, same). This quantification of some items was the main difference between version 1.0 and 1.1. These items accounted for the difference in scores between the two versions. Changes to other items resulted in only minor fluctuations. Some of the questions on DIRECT 1.1 required students to analyze simultaneous changes in variables, like voltage and resistance or current and voltage. Other questions required that students be proficient in their use of ratios.[31] Results indicated that students had difficulty with this analysis. The follow-up interviews indicated students' preference for and reliance on formulas.

Version 1.0 is more qualitative and seems to elicit the misconceptions more directly while version 1.1 is more quantitative and seems to elicit the students' mathematical abilities to some extent. If one is more interested in the conceptual understanding of circuits, version 1.0 and newer versions patterned after it would be the better alternative. However, if the students' mathematical abilities were of interest, then version 1.1 would be the appropriate choice.

We want to stress that DIRECT is not the end-all-be-all of tests. It simply provides another data point for instructors and researchers to use to evaluate the progress of students' understanding. No one instrument or study can provide definitive answers. Data regarding students' understanding should be considered like evidence of validity--requiring several measurements through different means to arrive at the final answer.

## Acknowledgements

The authors would like to acknowledge all the students and instructors who were involved in field testing DIRECT. Without their cooperation, this project would not have been possible. We would also like to thank the members of the independent panel of experts for their helpful and insightful feedback.

---

[1] D. Hestenes, M. Wells, and G. Swackhamer, "Force concept inventory," Phys. Teach. **30** (3), 141-58 (1992).

[2] R.J. Beichner, "Testing student interpretation of kinematics graphs," Am. J. Phys, 62 (8), 750-762 (1994).

[3] P. V. Engelhardt, "Examining students' understanding of electrical circuits through multiple-choice testing and interviews," unpublished doctoral dissertation, North Carolina State University (1997). The interested reader can read a more in-depth literature review in Chap. 2.

[4] M. Arnold and R. Millar, "Being constructive: An alternative approach to the teaching of introductory ideas in electricity," Int. J. Sci. Educ. **9** (5), 553-63. (1987); N. Fredette and J. Lochhead, "Student conceptions of simple circuits," Phys. Teach. **18** (3), 194-8 (1980); C. Kärrqvist, "Pupils are able," in *Proceedings of the Second International Seminar on Misconceptions and Educational Strategies in Science and Mathematics,* edited by J. Novak (Cornell University, Ithaca, NY, 1987), pp. 293-96; L.C. McDermott and E. H. van Zee, "Identifying and addressing student difficulties with electric xircuits," in *Aspects of Understanding Electricity: Proceedings of an International Workshop, Ludwigsburg, Germany*, edited by R. Duit, W. Jung, and C. von Rhöneck (Vertrieb Schmidt and Klaunig, Kiel, Germany, 1984), pp. 39-48; R. Osborne, "Children's ideas about electric current," New Zealand Science Teacher **29**, 12-19, (1981); D. M. Shipstone, "A study of children's understanding of electricity in simple DC circuits," Eur. J. Sci. Educ. **6** (2), 185-198 (1984).

[5] P. Licht and G. D. Thijs, "Method to trace coherence and persistence of preconceptions," Int. J. Sci. Educ. **12** (4), 403-416 (1990); R. Cohen, B. Eylon, and U. Ganiel, "Potential difference and current in simple electric circuits: A study of student's concepts," Am. J. Phys. **51** (5), 407-412 (1983); J. J. Dupin and S. Johsua, "Conceptions of French pupils concerning electric circuits: Structure and evolution," J. Res. Sci. Teach. **24** (9), 791-806 (1987).

[6] C. von Rhöneck and B. Völker, "Semantic structures describing the electric circuit before and after instruction," in *Aspects of Understanding Electricity: Proceedings of an International Workshop*, Ludwigsburg, Germany, edited by R. Duit, W. Jung, and C. von Rhöneck (Vertrieb Schmidt and Klaunig, Kiel, Germany, 1984), pp. 95-106; W. Jung, "Category questionnaires - The technique and some results," *ibid*., pp. 197-204; P. M. Heller and F. N. Finley, "Variable uses of alternative conceptions: A case study in current electricity," J. Res. Sci. Teach. **29** (3), 259-75 (1992).

[7] S. Johsua, "Students' interpretation of simple electrical diagrams," Eur. J. Sci. Educ. **6** (3), 271-275 (1984).

[8] M. Caillot, "Problem representations and problem-solving procedures in electricity," in *Aspects of Understanding Electricity*, *Proceedings of an International Workshop*, R. Duit, W. Jung, and C. von Rhöneck (Vertrieb Schmidt and Klaunig, Kiel, Germany, 1984), pp. 139-151; L. C. McDermott and P. S. Shaffer, "Research as a guide for curriculum development: An example from introductory electricity. Part I: Investigation of student understanding," Am. J. Phys. **60** (11), 994-1003 (1992).
[9] See McDermott and Shaffer, Ref. 7.

[10] R. Gott, "The place of electricity in the assessment of performance in science," in *Aspects of Understanding Electricity, Proceedings of an International Workshop,* edited by R. Duit, W. Jung, and C. von Rhöneck (Vertrieb Schmidt & Klaunig, Kiel, Germany, 1984), pp. 49-61.

[11] P. M. Heller and F. N. Finley, "Variable uses of alternative conceptions: A case study in current electricity," J. Res. Sci. Teach. **29** (3), 259-75 (1992); L.C. McDermott and P. S. Shaffer, "Research as a guide for curriculum development: An example from introductory electricity. Part I: Investigation of student understanding," Am. J. Phys. **60** (11), 994-1003 (1992).

[12] J. L. Closset, "Sequential reasoning in electricity," in *Research on Physics Education: Proceedings of the First International Workshop* (Editions du Centre National de la Recherche Scientifique, La Londe les Maures, France ,1984), pp. 313-319; D. M. Shipstone, "A study of children's understanding of electricity in simple DC circuits," Eur. J. Sci. Educ. **6** (2), 185-198 (1984).

[13] C. von Rhöneck and K. Grob, "Representation and problem-solving in basic electricity, predictors for successful learning" in *Proceedings of the Second International Seminar on Misconceptions and Educational Strategies in Science and Mathematics*, edited by J. D. Novak (Cornell University, Ithaca, NY, 1987), p. 564.

[14] J. M. Sebastià, "Cognitive mediators and interpretations of electric circuits," in *Proceedings of the Third International Seminar on Misconceptions and Educational Strategies in Science and Mathematics* (Misconceptions Trust, Ithaca, NY, 1993).

[15]R. Cohen, B. Eylon, and U. Ganiel, "Potential difference and current in simple electric circuits: A study of student's concepts," Am. J. Phys. **51** (5), 407-412 (1983); R. Millar and K. L. Beh, "Students' understanding of voltage in simple parallel electric circuits," Intl. J. Sci. Educ. **15** (4), 351-361 (1993); H. F. van Aalst, "The differentiation between connections in series and in parallel from cognitive mapping; Implications for teaching," in *Aspects of Understanding Electricity, Proceedings of an International Workshop*, edited by R. Duit, W. Jung, and C. von Rhöneck (Vertrieb Schmidt & Klaunig, Kiel, Germany, 1984), pp. 115-128.

[16] Ref. 14, van Aalst, p. 124.

[17] Arnold B. Arons, *Teaching Introductory Physics* (John Wiley & Sons, NY, 1997), p. 4.

[18] R. Cohen, B. Eylon, and U. Ganiel, "Potential difference and current in simple electric circuits: A study of student's concepts," Am. J. Phys. **51** (5), 407-412 (1983); J. J. Dupin and S. Johsua, "Conceptions of French pupils concerning electric circuits: Structure and evolution," J. Res. Sci. Teach. **24** (9), 791-806 (1987); K. Grob, V. L. Pollack, and C. von Rhöneck, "Computerized analysis of students' ability to process information in the area of basic electricity," in *Proceedings of the Research in Physics Learning: Theoretical Issues and Empirical Studies* , edited by F. G. Reinders Duit Hans Niedderer [xx one name? xx] (The Institute for Science Education at the University of Kiel, Kiel, Germany, 1991), pp. 296-309; S.M. Lea, B.A. Thacker, E. Kim, and K. M. Miller, "Computer-assisted assessment of student understanding in physics," Computers in Physics **8** (1), 122-127 (1994); P. Licht and G. D. Thijs, "Method to trace coherence and persistence of preconceptions," Intl. J. Sci. Educ. **12** (4), 403-416 (1990); R. Millar and T. King, "Students' understanding of voltage in simple series electric circuits," Intl. J. Sci. Educ. **15** (3), 339-349 (1993); D. R. Sokoloff, "RealTime [xx one word or 2? xx] physics electricity: Active learning of electric circuit concepts

using microcomputer-based current and voltage probes," to be published in La Fisica Nella Scuola (1994); [xx 1994? xx] M. S. Steinberg and C. L. Wainwright, "Using models to teach electricity: The CASTLE project," Phys. Teach. **31** (9), 353-57 (1993).

[19] A. H. Johnstone and A.R. Mughol, "The concept of electrical resistance," Phys. Educ. **13**, 46-49 (1978).

[20] D. M. Shipstone, C. von Rhöneck, W. Jung, C. Kärrqvist, J. J. Dupin, S. Johsua, and P. Licht, "A study of students' understanding of electricity in five European countries," Intl. J. Sci. Educ. **10** (3), 303-16 (1988).

[21] P. S. Shaffer and L. C. McDermott, "Research as a guide for curriculum development: An example from introductory electricity. Part II: Design of instructional strategies," Am. J. Phys. **60** (11), 1003-13 (1992).

[22] L. C. McDermott and P. S. Shaffer, "Research as a guide for curriculum development: An example from introductory electricity. Part I: Investigation of student understanding," Am. J. Phys. **60** (11), 999-1000 (1992).

[23] D. Psillos, P. Koumaras, and O. Valassiades, "Pupils' representations of electric current before, during and after instruction on DC circuits," Res. Sci. and Technological Educ. **5** (2), 193 (1987).

[24] L. C. McDermott et al., *Physics by Inquiry* (John Wiley & Sons, NY, 1996), Vol. II.

[25] R. A. Serway and J. S. Faughn, *College Physics* (Saunders College Publishing, NY, 1985).

[26] R. Chabay and B. Sherwood, *Electric & Magnetic Interactions* (John Wiley & Sons, NY, 1995).

[27] R. Cohen, B. Eylon, and U. Ganiel, "Potential difference and current in simple electric circuits: A study of student's concepts," Am. J. Phys. **51** (5), 407-412 (1983).

[28] J. W. Best, *Research in Education* (Prentice-Hall, Englewood Cliffs, NJ, 1981), 4th ed., p. 14.

[29] DIRECT version 1.1 is available from <http://www.ncsu.edu/PER>.

[30] These results and the accompanying tables are available at <http://www.ncsu.edu/PER>. Files containing both versions of the test are also available.

[31] A. B. Arons, *A Guide to Introductory Physics Teaching* (John Wiley & Sons, NY, 1990), pp. 3-6.

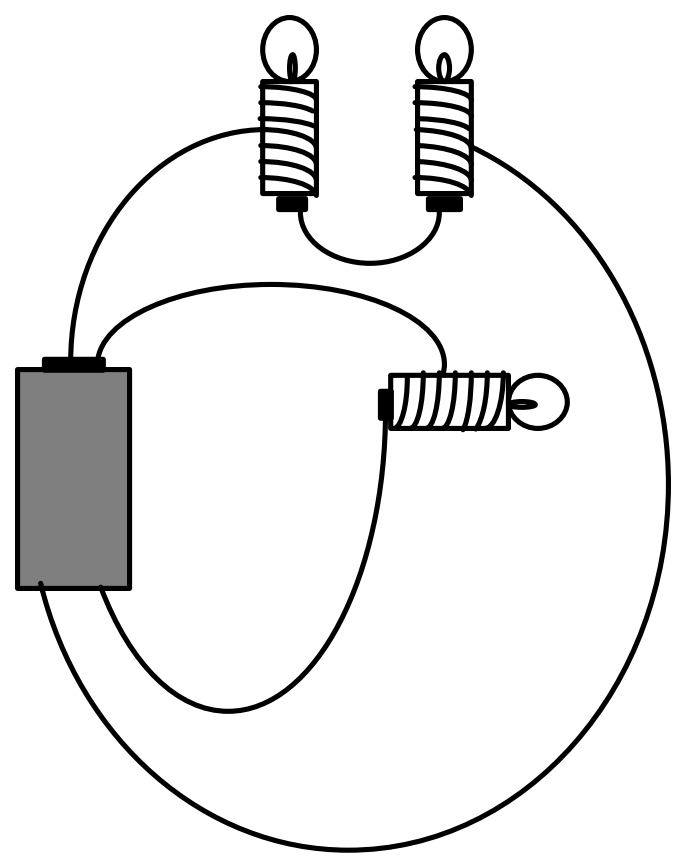

# **D**etermining and **I**nterpreting **R**esistive **E**lectric Circuits **C**oncepts **T**est Version 1.0

## *Instructions*

**Wait until you are told to begin**, then turn to the next page and begin working. Answer each question as accurately as you can. There is only one correct answer for each item. Feel free to use a calculator and scratch paper if you wish.

Use a #2 pencil to **record you answers** on the computer sheet, but please **do not write in the test booklet**.

You will have approximately one hour to complete the test. If you finish early, check your work before handing in both the answer sheet and the test booklet.

## *Additional comments about the test*

All light bulbs, resistors, and batteries should be considered identical unless you are told otherwise. The battery is to be assumed ideal, that is to say, the internal resistance of the battery is negligible. In addition, assume the wires have negligible resistance. Below is a key to the symbols used on this test. Study them carefully before you begin the test.

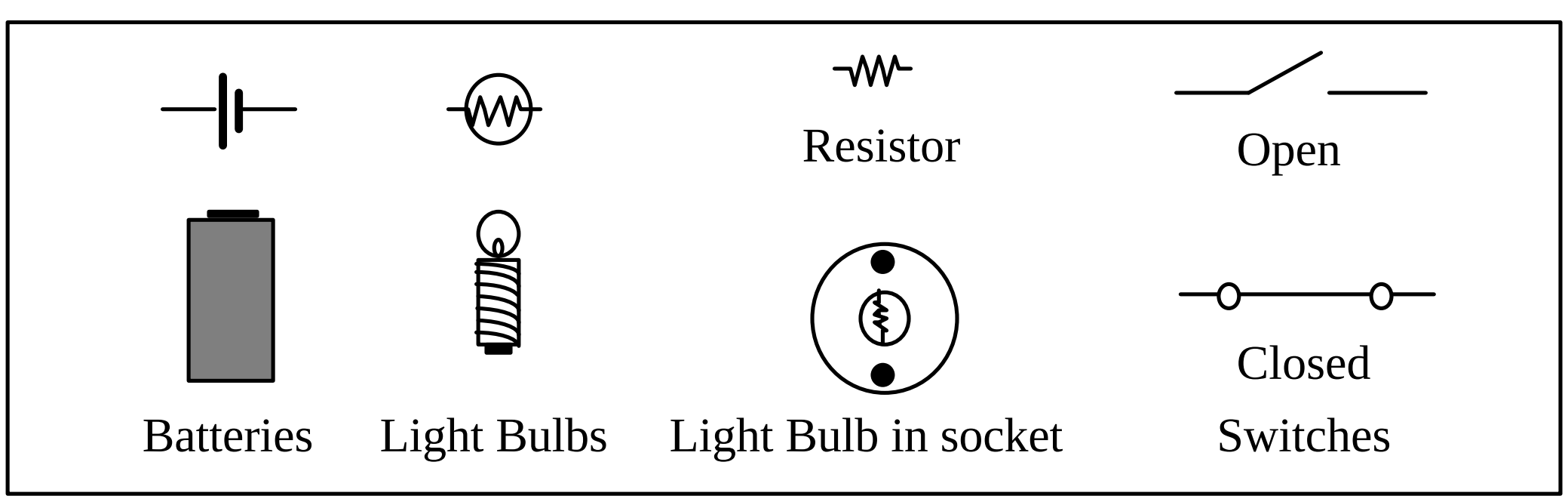

North Carolina State University
Department of Physics
Raleigh, NC 27695-8202

1) Are charges used up in a light bulb, being converted to light?

(A) Yes, charges moving through the filament produce “friction” which heats up the filament and produces light.

(B) Yes, charges are emitted.

(C) No, charge is conserved. It is simply converted to another form such as heat and light.

(D) No, charge is conserved. Charges moving through the filament produce “friction” which heats up the filament and produces light.

2) How does the power delivered to resistor A change when resistor B is added as shown in circuits 1 and 2 respectively?

(A) Increases
(B) Decreases
(C) Stays the same

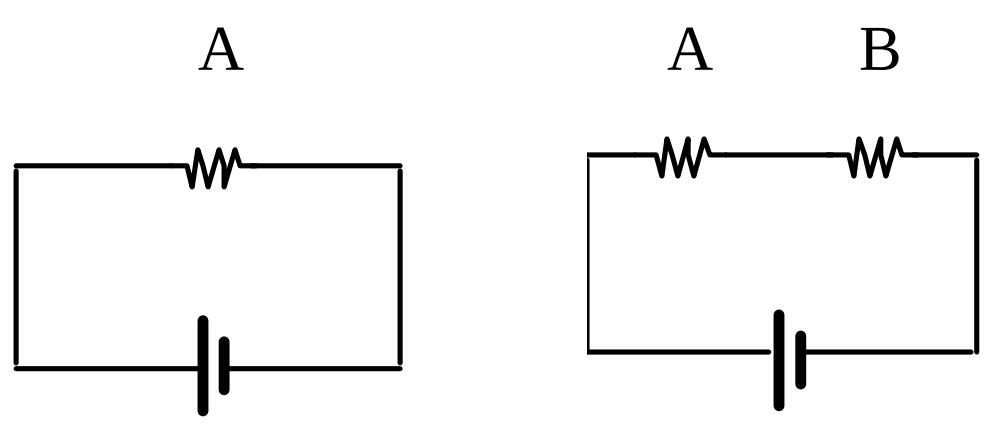

Circuit 1 Circuit 2

3) Consider the circuits shown below. Which circuit or circuits have the greatest energy delivered to it per second?

(A) Circuit 1
(B) Circuit 2
(C) Circuit 3
(D) Circuit 1 = Circuit 2
(E) Circuit 2 = Circuit 3

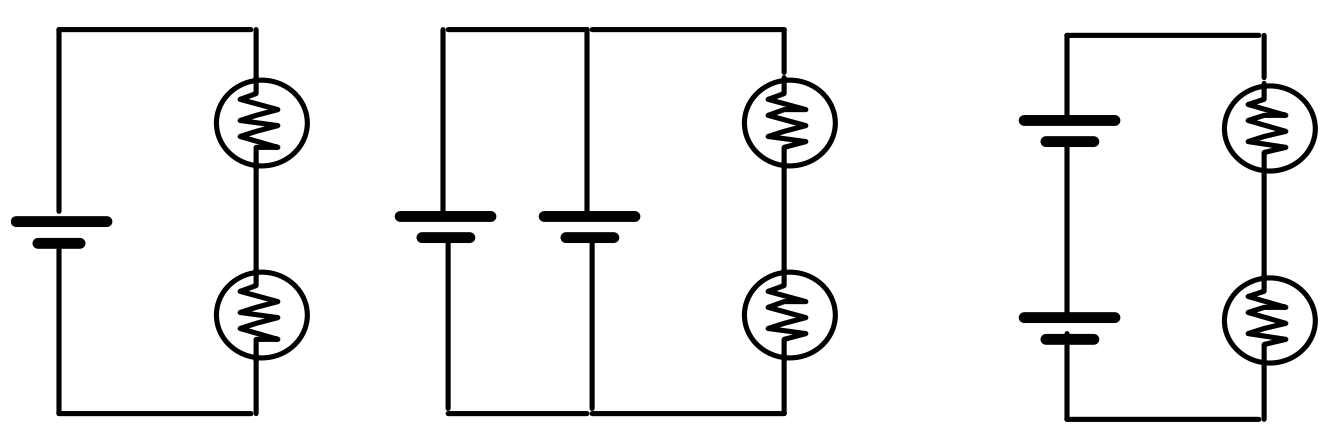

Circuit 1 Circuit 2 Circuit 3

4) Consider the following circuits.

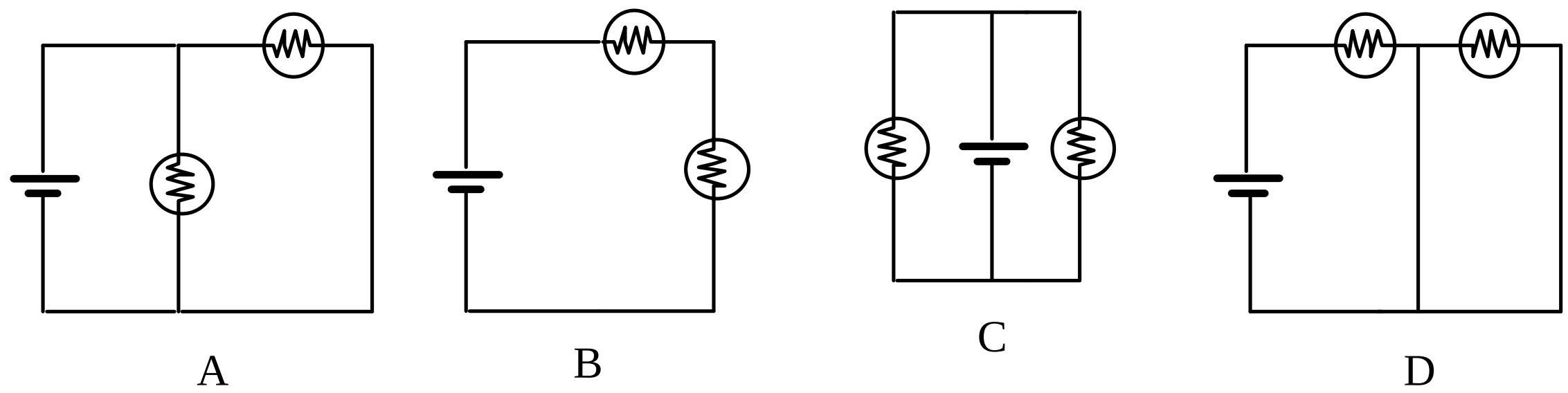

Which circuit(s) above represent(s) a circuit consisting of two light bulbs in parallel with a battery?

(A) A
(B) B
(C) C
(D) A and C
(E) A, C, and D

5) Compare the resistance of branch 1 with that of branch 2. A branch is a section of a circuit. Which has the least resistance?

(A) Branch 1
(B) Branch 2
(C) Neither, they are the same

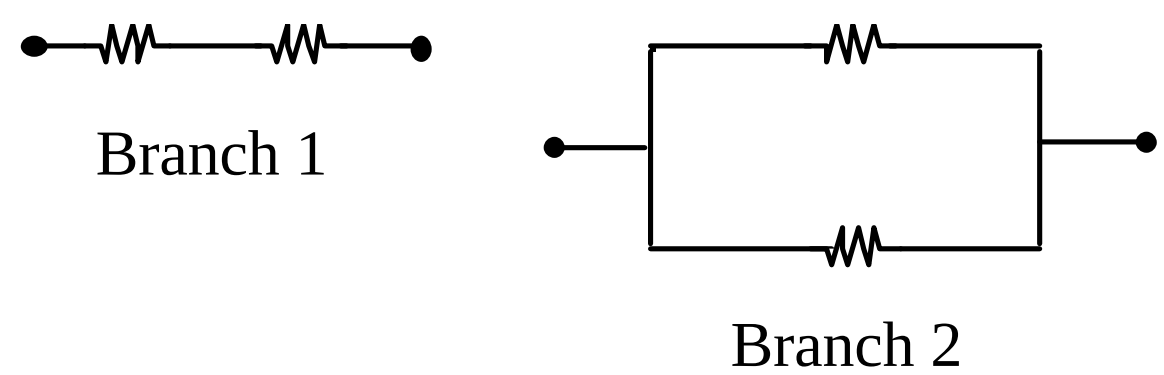

6) Rank the potential difference between points 1 and 2, points 3 and 4, and points 4 and 5 in the circuit shown below from highest to lowest.

(A) 1 and 2; 3 and 4; 4 and 5
(B) 1 and 2; 4 and 5; 3 and 4
(C) 3 and 4; 4 and 5; 1 and 2
(D) 3 and 4 = 4 and 5; 1 and 2
(E) 1 and 2; 3 and 4 = 4 and 5

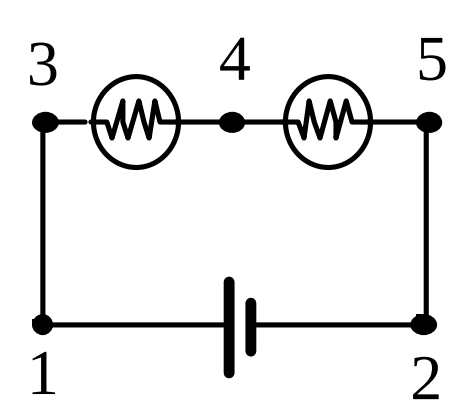

7) Compare the brightness of the bulb in circuit 1 with that in circuit 2. Which bulb is brighter?

(A) Bulb in circuit 1
(B) Bulb in circuit 2
(C) Neither, they are the same

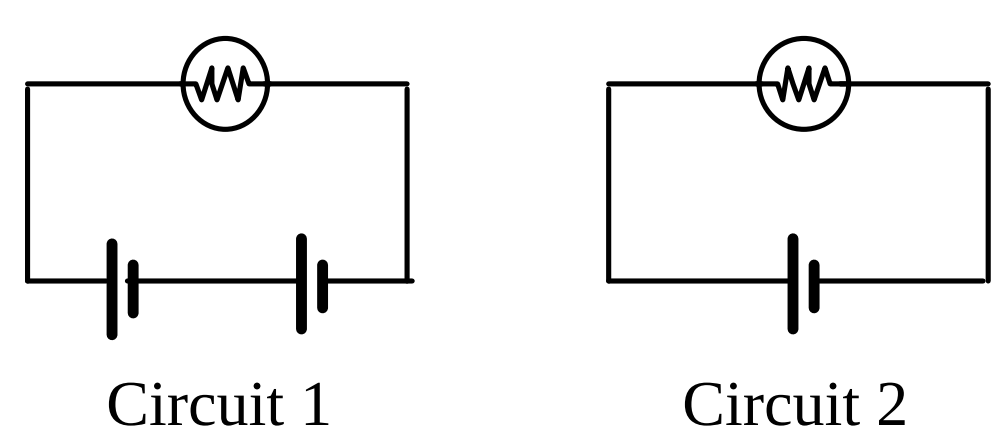

8) Compare the current at point 1 with the current at point 2. Which point has the larger current?

(A) Point 1
(B) Point 2
(C) Neither, they are the same

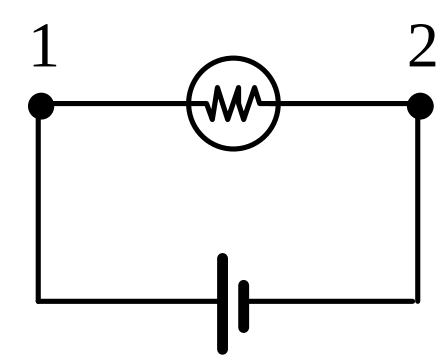

9) Which circuit(s) will light the bulb?

(A) A
(B) C
(C) D
(D) A and C
(E) B and D

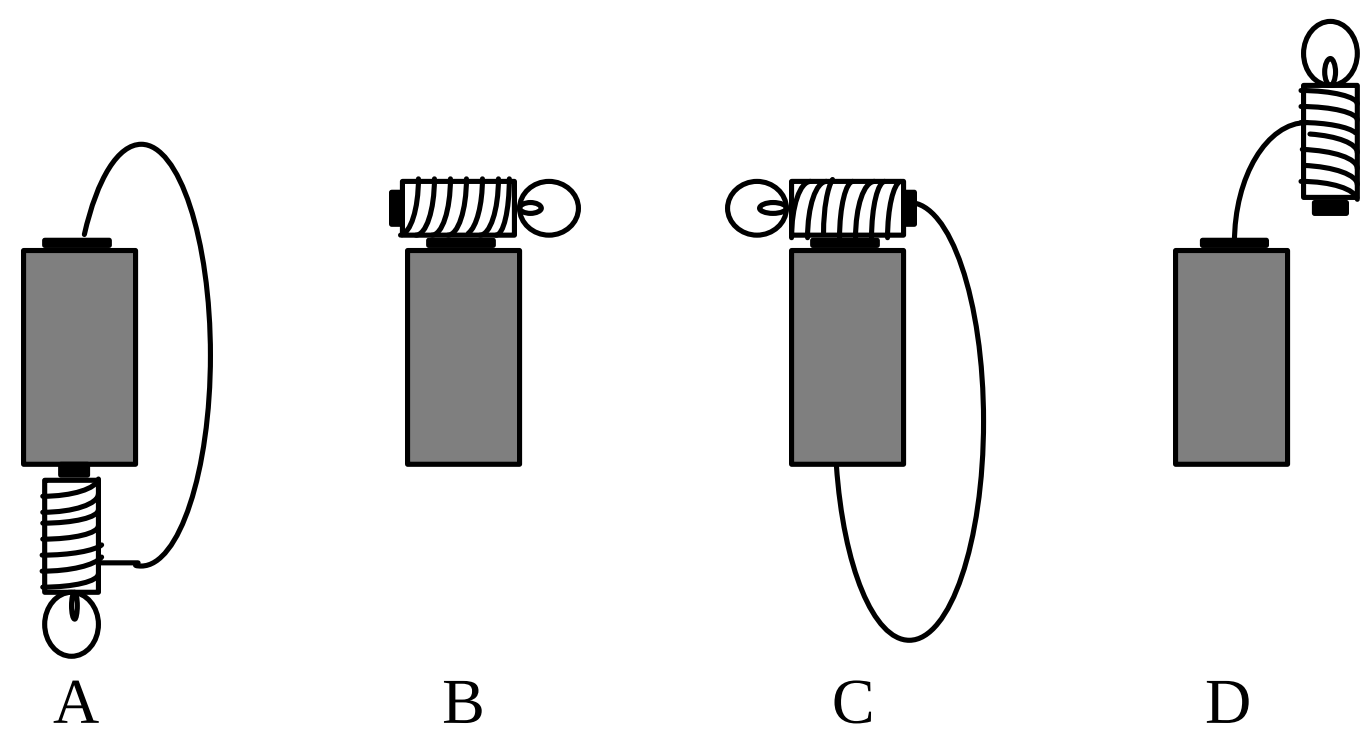

10) Compare the brightness of bulbs A and B in circuit 1 with the brightness of bulb C in circuit 2. Which bulb or bulbs are the brightest?

(A) A
(B) B
(C) C
(D) A = B
(E) A = C

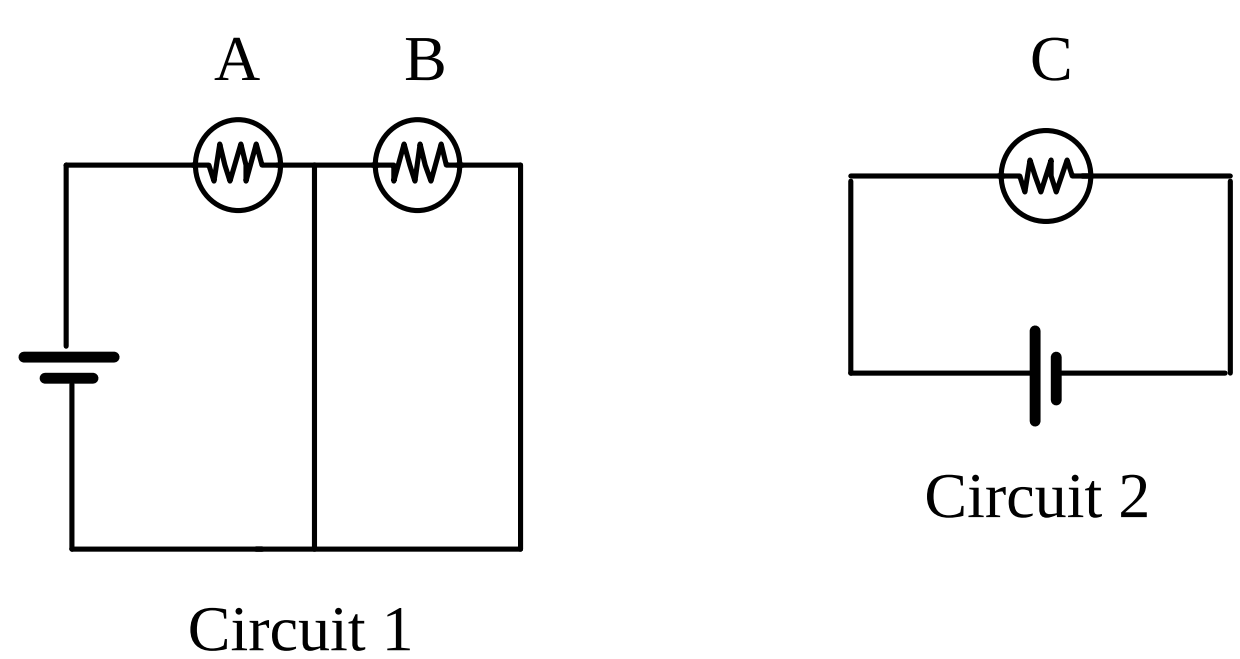

11) Why do the lights in your home come on almost instantaneously?

(A) Charges are already in the wire. When the circuit is completed, there is a rapid rearrangement of surface charges in the circuit.

(B) Charges store energy. When the circuit is completed, the energy is released.

(C) Charges in the wire travel very fast.

(D) The circuits in a home are wired in parallel. Thus, a current is already flowing.

12) Consider the power delivered to each of the resistors shown in the circuits below. Which circuit or circuits have the least power delivered to it?

(A) Circuit 1
(B) Circuit 2
(C) Circuit 3
(D) Circuit 1 = Circuit 2
(E) Circuit 1 = Circuit 3

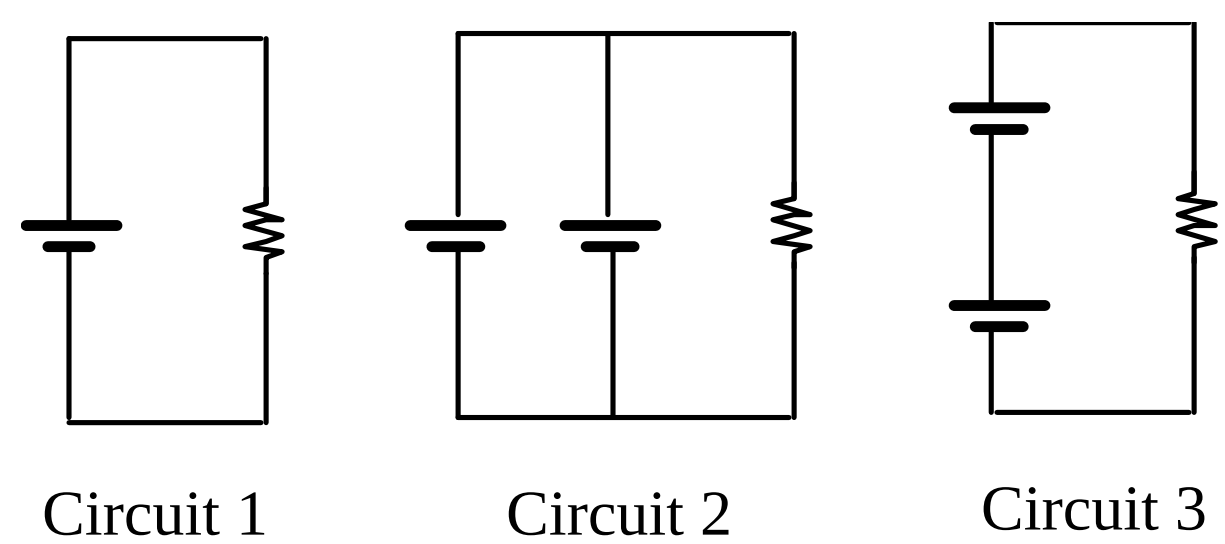

13) Which schematic diagram best represents the realistic circuit shown below?

(A) A
(B) B
(C) C
(D) D
(E) None of the above

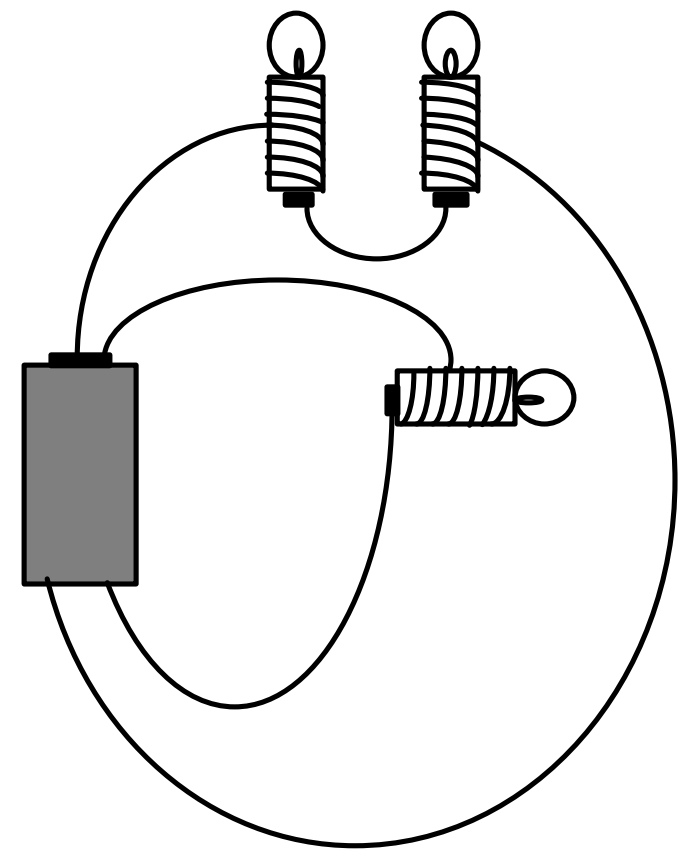

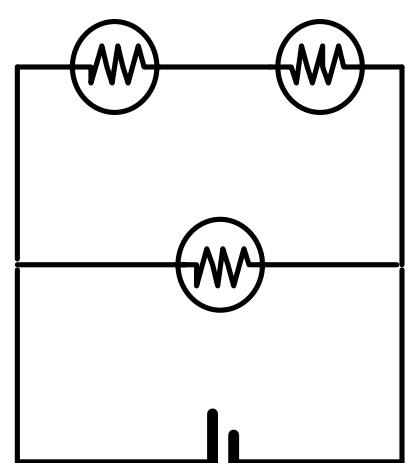

A

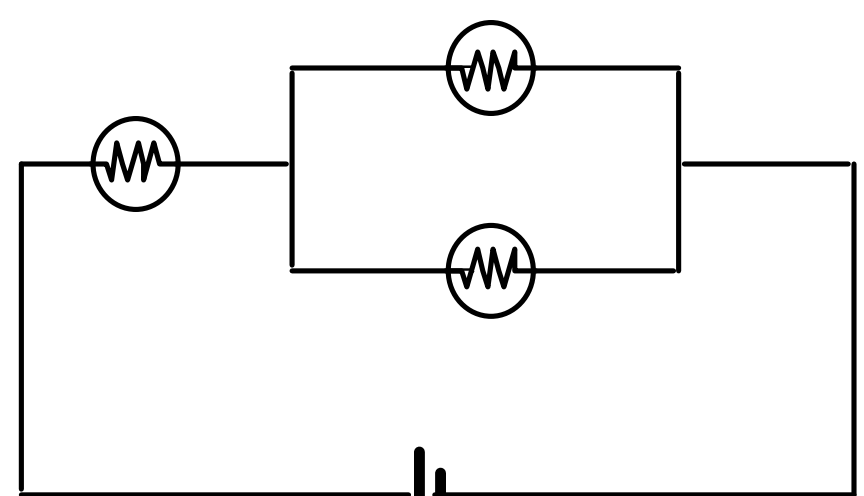

B

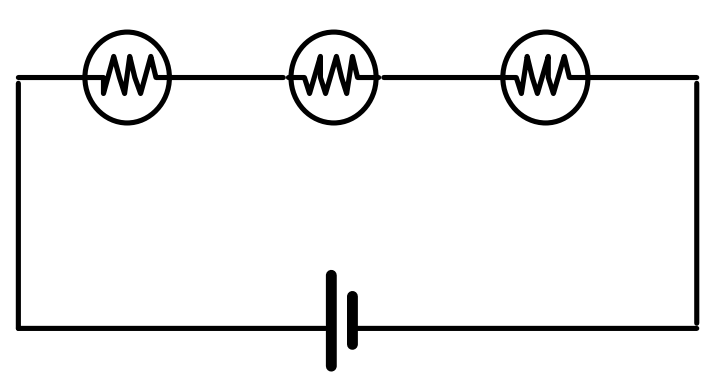

C

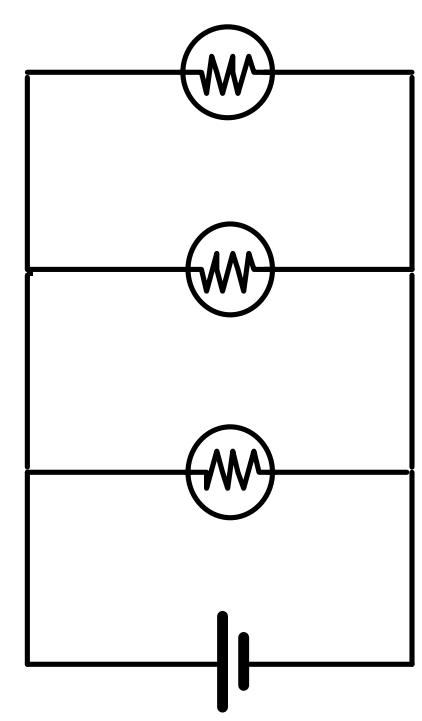

D

14) How does the resistance between the endpoints change when the switch is closed?

(A) Increases
(B) Decreases
(C) Stays the same

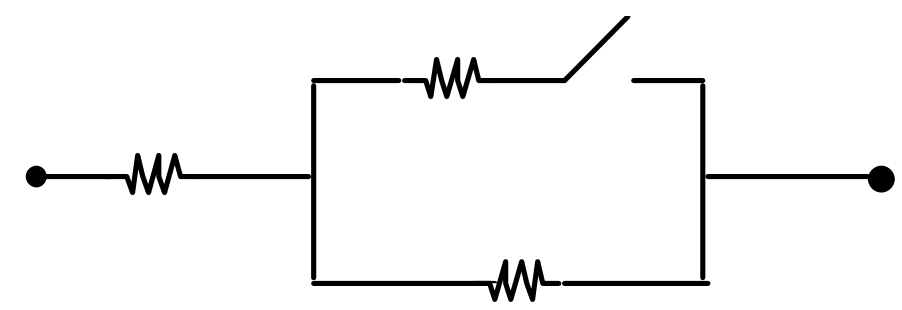

15) What happens to the potential difference between points 1 and 2 if bulb A is removed?

(A) Increases
(B) Decreases
(C) Stays the same

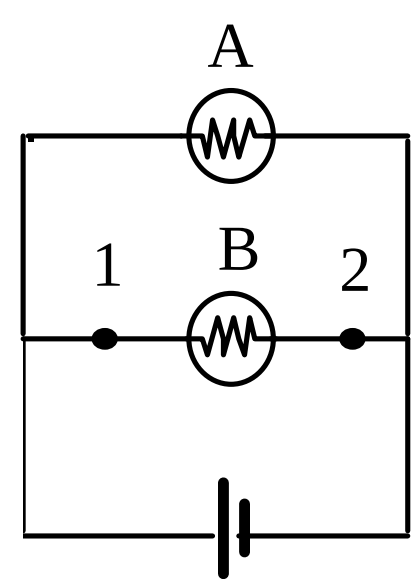

16) Compare the brightness of bulb A in circuit 1 with bulb A in circuit 2. Which bulb is dimmer?

(A) Bulb A in circuit 1
(B) Bulb A in circuit 2
(C) Neither, they are the same

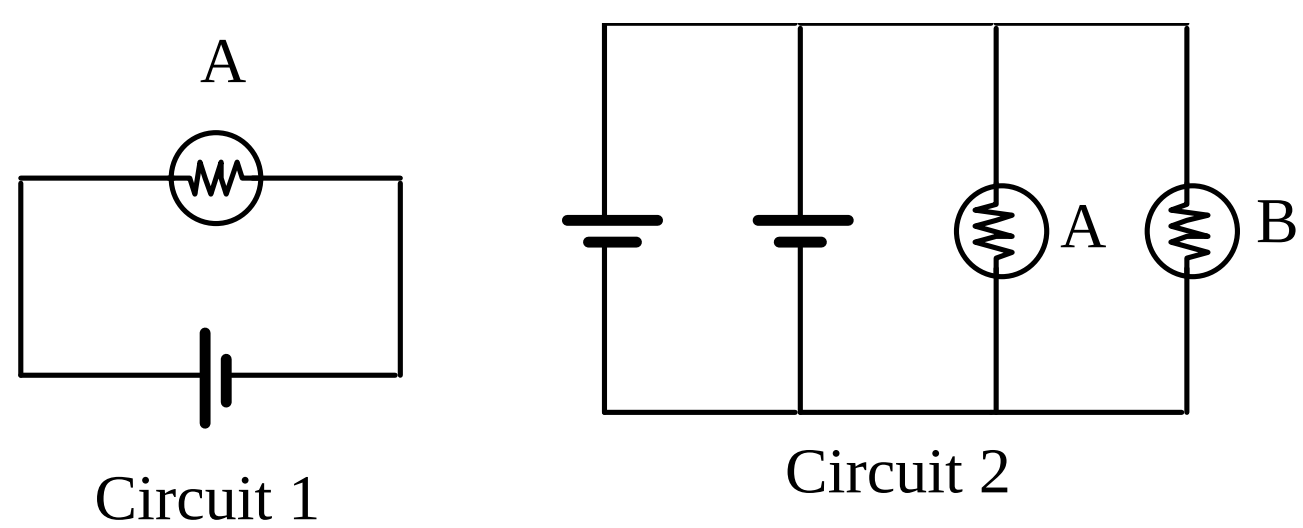

17) Rank the currents at points 1, 2, 3, 4, 5, and 6 from highest to lowest.

(A) 5, 1, 3, 2, 4, 6
(B) 5, 3, 1, 4, 2, 6
(C) 5 =6, 3 = 4, 1 =2
(D) 5 = 6, 1 = 2 = 3 = 4
(E) 1 = 2 = 3 = 4 = 5 = 6

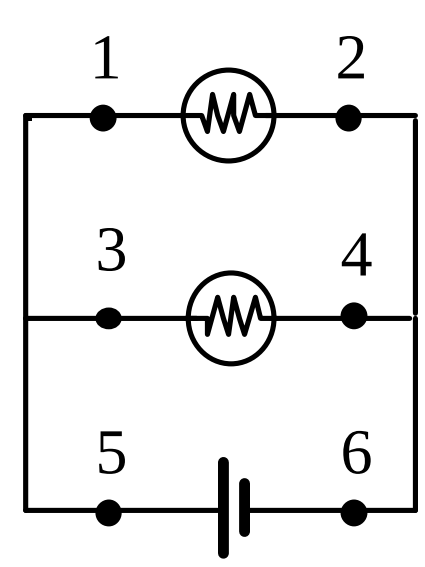

18) Which circuit(s) will light the bulb?

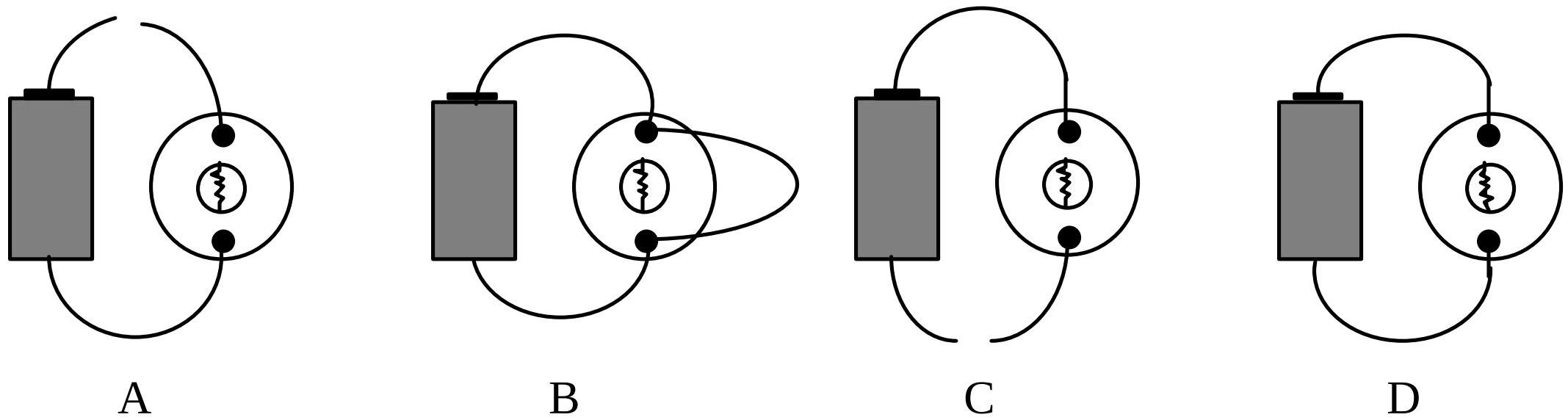

(A) A
(B) B
(C) D
(D) B and D
(E) A and C

19) What happens to the brightness of bulbs A and B when a wire is connected between points 1 and 2?

(A) Increases
(B) Decreases
(C) Stays the same
(D) A becomes brighter than B
(E) Neither bulb will light

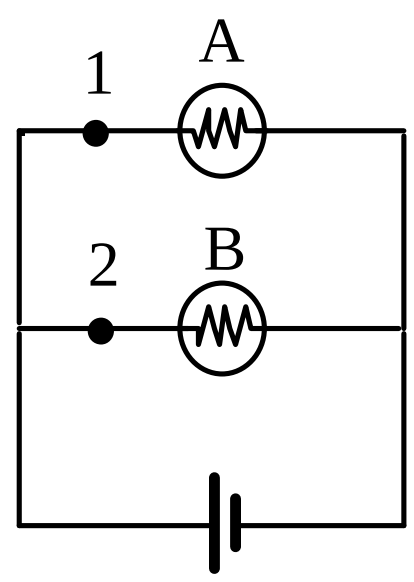

20) Is the electric field zero or non-zero inside the tungsten bulb filament?

(A) Zero because the filament is a conductor.
(B) Zero because there is a current flowing.
(C) Non-zero because the circuit is complete and a current is flowing.
(D) Non-zero because there are charges on the surface of the filament.

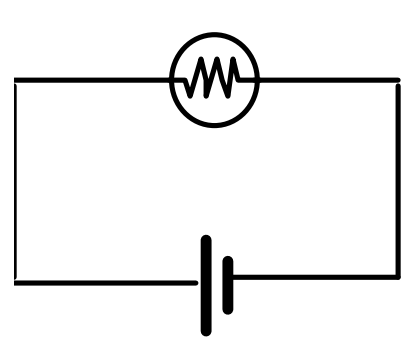

21) Compare the energy delivered per second to the light bulb in circuit 1 with the energy delivered per second to the light bulbs in circuit 2. Which bulb or bulbs have the least energy delivered to it per second?

(A) A
(B) B
(C) C
(D) B = C
(E) A = B = C

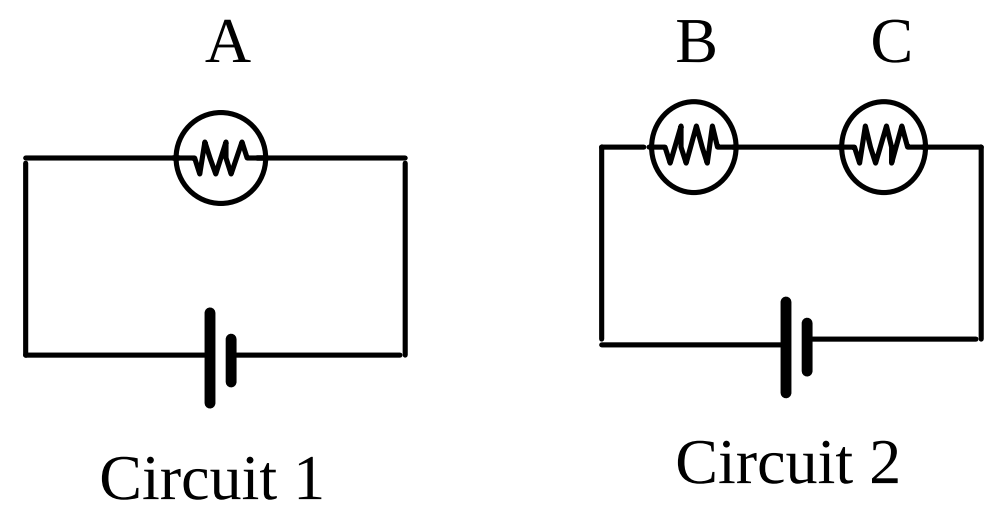

22) Which realistic circuit(s) represent(s) the schematic diagram shown below?

(A) B
(B) C
(C) D
(D) A and B
(E) C and D

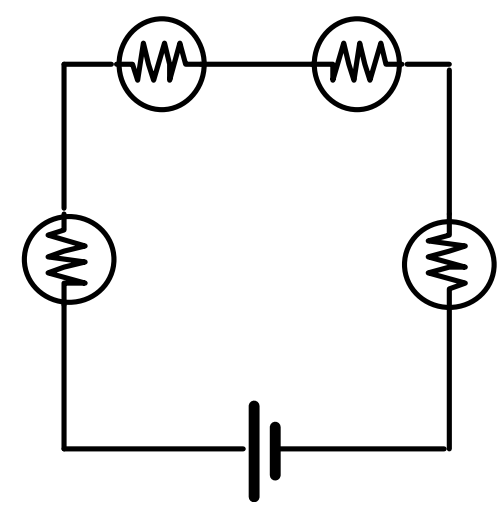

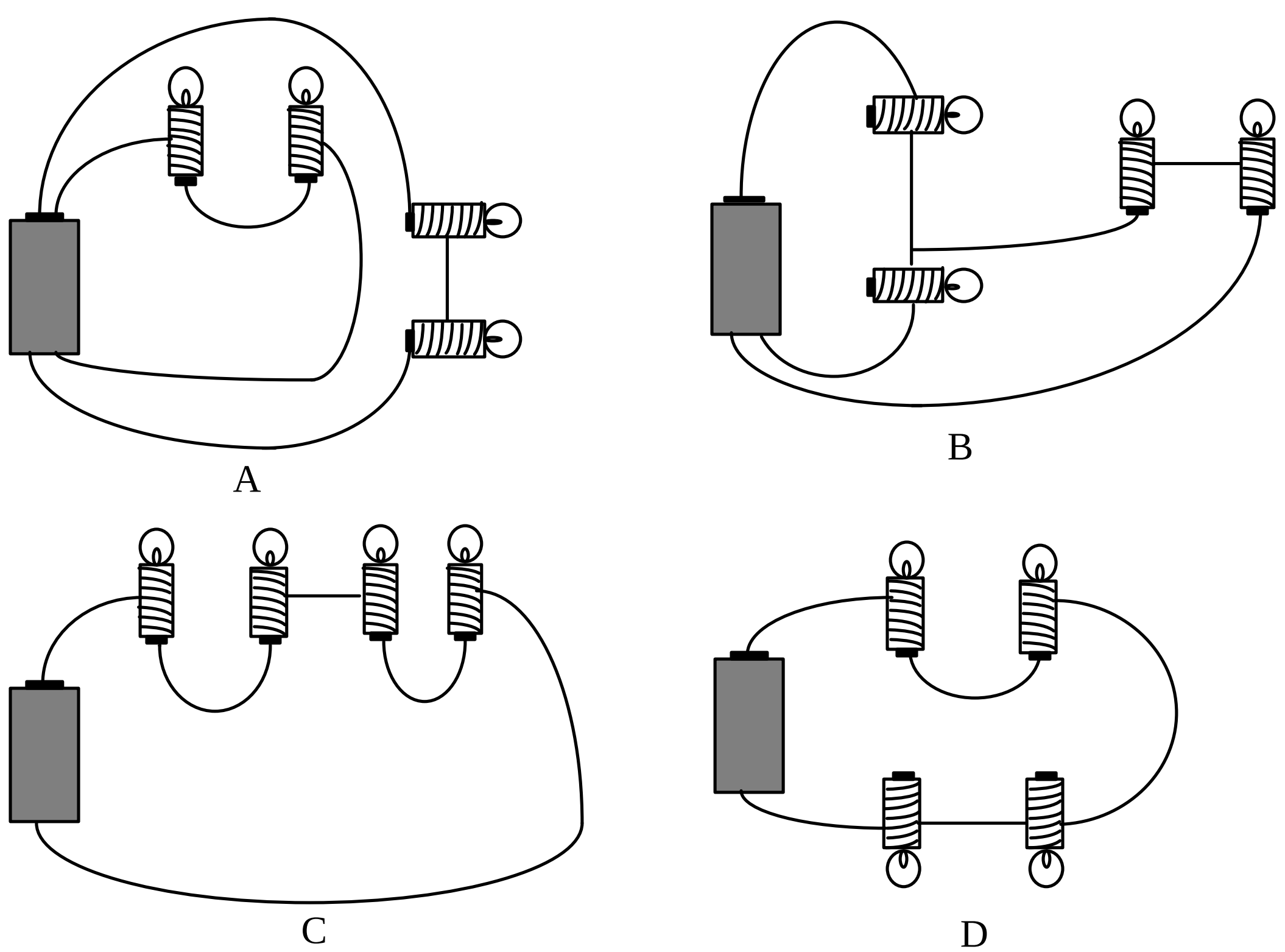

23) Immediately after the switch is opened, what happens to the resistance of the bulb?

(A) The resistance increases.
(B) The resistance decreases.
(C) The resistance stays the same.
(D) The resistance goes to zero.

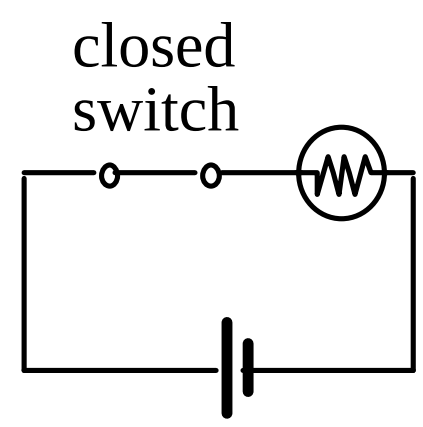

24) If you double the current through a battery, is the potential difference across a battery doubled?

(A) Yes, because Ohm's law says $V = IR$.
(B) Yes, because as you increase the resistance, you increase the potential difference.
(C) No, because as you double the current, you reduce the potential difference by half.
(D) No, because the potential difference is a property of the battery.
(E) No, because the potential difference is a property of everything in the circuit.

25) Compare the brightness of bulb A in circuit 1 with bulb A in circuit 2. Which bulb is brighter?

(A) Bulb A in circuit 1
(B) Bulb A in circuit 2
(C) Neither, they are the same

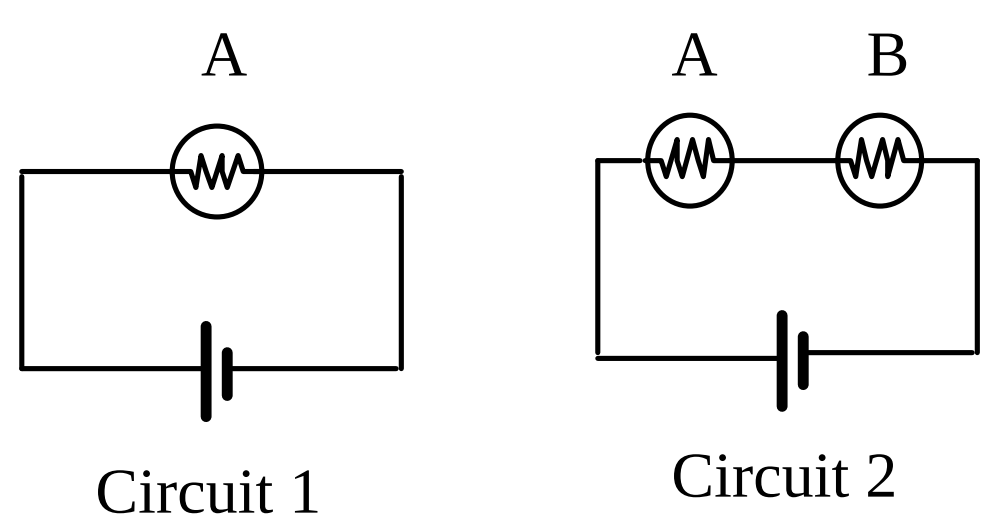

26) If you increase the resistance C, what happens to the brightness of bulbs A and B?

(A) A stays the same, B dims
(B) A dims, B stays the same
(C) A and B increase
(D) A and B decrease
(E) A and B remain the same

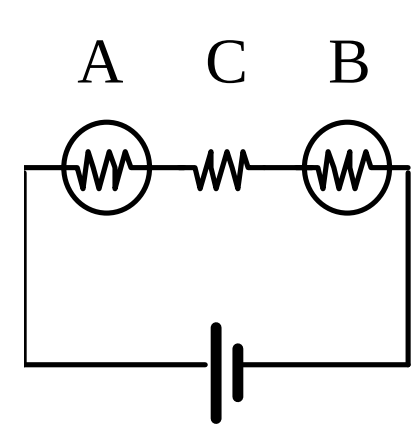

27) Will all the bulbs be the same brightness?

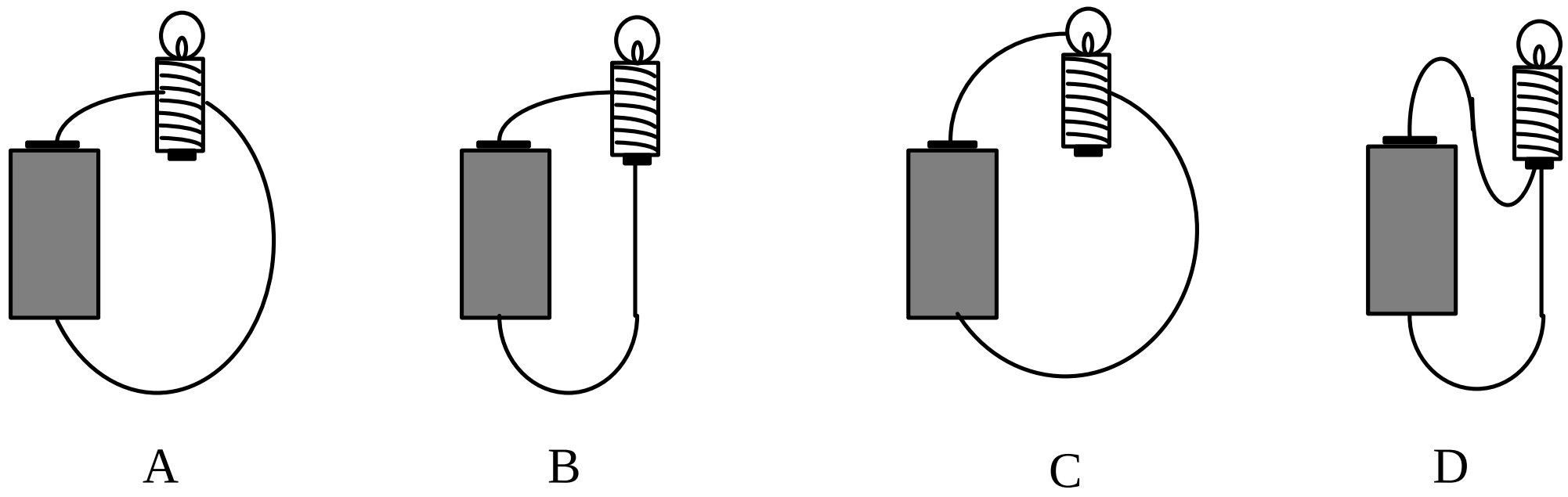

(A) Yes, because they all have the same type of circuit wiring.
(B) No, because only B will light. The connections to A, C, and D are not correct.
(C) No, because only D will light. D is the only complete circuit.
(D) No, C will not light but A, B, and D will.

28) What is the potential difference between points A and B?

(A) 0 V
(B) 3 V
(C) 6 V
(D) 12 V

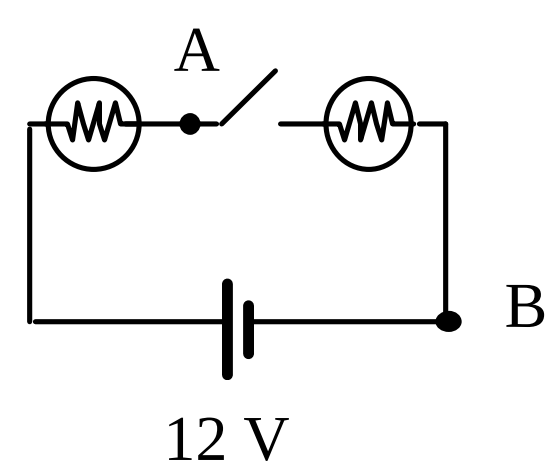

29) What happens to the brightness of bulbs A and B when the switch is closed?

(A) A stays the same, B dims
(B) A brighter, B dims
(C) A and B increase
(D) A and B decrease
(E) A and B remain the same

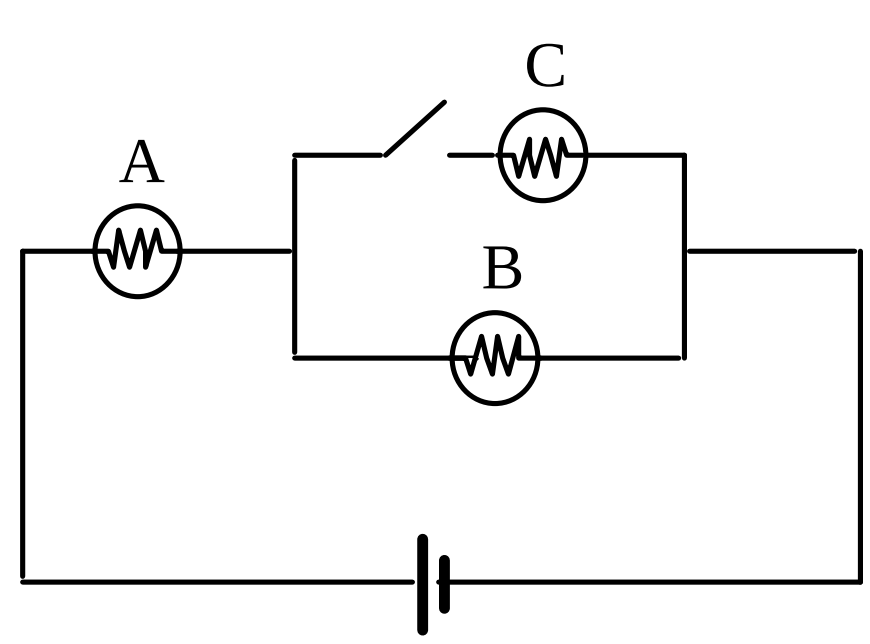

Table I: Objectives for DIRECT and results.

| Objective | Question Number | Average Percentage Correct | |
|---|---|---|---|
| | | v. 1.0 | v. 1.1 |
| **Physical Aspects of DC electric circuits (objectives 1-5)** | | 56 | 52 |
| 1) Identify and explain a short circuit (more current follows the path of lesser resistance). | 10, 19, 27 | 56 | 56 |
| 2) Understand the functional two-endedness [xx awkward xx[ of circuit elements (elements have two possible points with which to make a connection).<br>3) Identify a complete circuit and understand the necessity of a complete circuit for current to flow in the steady state (some charges are in motion but their velocities at any location are not changing and there is no accumulation of excess charge anywhere in the circuit). | 9, 18 | 54 | 59 |
| **Objectives 1-3 combined** | 27 | 68 | 73 |
| 4) Apply the concept of resistance (the hindrance to the flow of charges in a circuit) including that resistance is a property of the object (geometry of object and type of material with which the object is composed) and that in series the resistance increases as more elements are added and in parallel the resistance decreases as more elements are added. | 5, 14, 23 | 59 | 40 |
| 5) Interpret pictures and diagrams of a variety of circuits including series, parallel, and combinations of the two. | 4, 13, 22 | 55 | 54 |
| **Circuit layout (objectives 1-3,5)** | | 55 | 56 |
| **Energy (objectives 6-7)** | | 42 | 31 |
| 6) Apply the concept of power (work done per unit time) to a variety of circuits. | 2, 12 | 37 | 28* |
| 7) Apply a conceptual understanding of conservation of energy including Kirchhoff's loop rule (? V=0 around a closed loop) and the battery as a source of energy. | 3, 21 | 47 | 49 |
| **Current (objectives 8-9)** | | 44 | 44 |
| 8) Understand and apply conservation of current (conservation of charge in the steady state) to a variety of circuits. | 8, 17 | 62 | 59 |
| 9) Explain the microscopic aspects of current flow in a circuit through the use of electrostatic terms such as electric field, potential differences, and the interaction of forces on charged particles. | 1, 11, 20 | 31 | 19 |
| **Potential difference (Voltage) (objectives 10-11)** | | 46 | 35 |
| 10) Apply the knowledge that the amount of current is influenced by the potential difference maintained by the battery and resistance in the circuit. | 7, 16, 25 | 60 | 38 |
| 11) Apply the concept of pot. diff. to a variety of circuits including the knowledge that the pot. diff. in a series circuit sums while in a parallel circuit it remains the same. | 6, 15, 24, 28, 29 | 37 | 34 |
| **Current and Voltage (objectives 8 and 11)** | 26 | 45 | 40 |

Table II: Statistical results for DIRECT.

| **Statistic** | **Value for version 1.0** | **Value for version 1.1** | **Ideal value** | **What it measures** |
|---|---|---|---|---|
| N | 1135 | 692 | Large to reduce sampling error | Number of students taking the test |
| Overall Mean | 48 ± 0.45% | 41 ± 0.55% | 50% for maximum spread of scores | |
| University Mean | 52 ± 0.56% | 44 ± 0.69% | | |
| High school Mean | 41 ± 0.65% | 36 ± 0.79% | | |
| Standard error of the mean | 0.45 | 0.55 | As close to zero as possible | Uncertainty in the mean |
| Overall Range | 14 - 97% | 3.4 - 90% | 0 - 100 | |
| University Range | 21 - 97% | 10 - 90% | 0 - 100 | |
| High school Range | 14 - 90% | 3.4 - 76% | 0 - 100 | |
| Kuder-Richardson 20 (KR-20) or Reliability | 0.71 | 0.70 | ≥0.70 for group measurement | Internal consistency of the instrument |
| Average Point-biserial correlation | 0.33 | 0.32 | ≥ 0.20 | Reliability of a single item on the test |
| Average discrimination index | 0.26 | 0.23 | ≥ 0.30 | Ability of a single item to differentiate between students scoring well on the test and students scoring poorly |
| Average difficulty index | 0.49 | 0.41 | 0.40 - 0.60 | Proportion of students in the sample who chose the correct response |

Table III. Results for DIRECT version 1.0 for each question. The fraction choosing the correct answer is in bold. A detailed breakdown by level (high school and university) is available on the web.[25]

| | Fraction picking letter choice | | | | | | | |
|---|---|---|---|---|---|---|---|---|
| Question | A | B | C | D | E | Correlation | Discrimination | Difficulty |
| 1 | 0.20 | 0.03 | 0.32 | **0.46** | 0.00 | 0.33 | 0.29 | 0.46 |
| 2 | 0.13 | **0.55** | 0.32 | 0.00 | 0.00 | 0.30 | 0.21 | 0.55 |
| 3 | 0.04 | 0.31 | **0.42** | 0.05 | 0.18 | 0.35 | 0.26 | 0.42 |
| 4 | 0.08 | 0.03 | 0.30 | **0.43** | 0.16 | 0.38 | 0.33 | 0.43 |
| 5 | 0.10 | **0.78** | 0.11 | 0.00 | 0.00 | 0.37 | 0.27 | 0.78 |
| 6 | 0.15 | 0.06 | 0.06 | 0.15 | **0.58** | 0.41 | 0.36 | 0.58 |
| 7 | **0.63** | 0.10 | 0.27 | 0.00 | 0.00 | 0.30 | 0.22 | 0.63 |
| 8 | 0.17 | 0.03 | **0.80** | 0.00 | 0.00 | 0.37 | 0.26 | 0.80 |
| 9 | 0.12 | 0.04 | 0.03 | **0.79** | 0.01 | 0.32 | 0.24 | 0.79 |
| 10 | 0.02 | 0.01 | 0.53 | 0.11 | **0.33** | 0.14 | 0.09 | 0.33 |
| 11 | **0.33** | 0.11 | 0.21 | 0.36 | 0.00 | 0.18 | 0.10 | 0.33 |
| 12 | 0.37 | 0.16 | 0.13 | **0.19** | 0.14 | 0.39 | 0.22 | 0.19 |
| 13 | **0.89** | 0.04 | 0.01 | 0.01 | 0.04 | 0.29 | 0.15 | 0.89 |
| 14 | 0.30 | **0.57** | 0.13 | 0.00 | 0.00 | 0.40 | 0.32 | 0.57 |
| 15 | 0.36 | 0.12 | **0.52** | 0.00 | 0.00 | 0.31 | 0.24 | 0.52 |
| 16 | 0.24 | 0.26 | **0.49** | 0.00 | 0.00 | 0.17 | 0.09 | 0.49 |
| 17 | 0.02 | 0.11 | 0.13 | **0.44** | 0.30 | 0.44 | 0.35 | 0.44 |
| 18 | 0.00 | 0.02 | **0.28** | 0.68 | 0.01 | 0.32 | 0.21 | 0.28 |
| 19 | 0.03 | 0.13 | **0.67** | 0.08 | 0.08 | 0.33 | 0.26 | 0.67 |
| 20 | 0.14 | 0.08 | 0.63 | **0.15** | 0.00 | 0.07 | 0.01 | 0.15 |
| 21 | 0.07 | 0.04 | 0.23 | **0.51** | 0.16 | 0.43 | 0.34 | 0.51 |
| 22 | 0.02 | **0.32** | 0.18 | 0.04 | 0.44 | 0.33 | 0.29 | 0.32 |
| 23 | 0.09 | 0.11 | **0.41** | 0.39 | 0.00 | 0.35 | 0.28 | 0.41 |
| 24 | 0.47 | 0.06 | 0.16 | **0.25** | 0.05 | 0.46 | 0.29 | 0.25 |
| 25 | **0.69** | 0.04 | 0.27 | 0.01 | 0.00 | 0.36 | 0.27 | 0.69 |
| 26 | 0.37 | 0.05 | 0.07 | **0.45** | 0.06 | 0.51 | 0.43 | 0.45 |
| 27 | 0.06 | **0.68** | 0.10 | 0.15 | 0.00 | 0.42 | 0.38 | 0.68 |
| 28 | 0.56 | 0.03 | 0.19 | **0.21** | 0.00 | 0.07 | 0.00 | 0.21 |
| 29 | 0.31 | **0.31** | 0.16 | 0.10 | 0.10 | 0.38 | 0.28 | 0.31 |
| | | | | | Average | 0.33 | 0.24 | 0.49 |

Table IV: Results for DIRECT version 1.1 for each question. Fraction choosing the correct answer is in bold. A detailed breakdown by level (high school and university) is available on the web.[25]

| | Fraction picking letter choice | | | | | | | |
|---|---|---|---|---|---|---|---|---|
| Question | A | B | C | D | E | Correlation | Discrimination | Difficulty |
| 1 | 0.13 | 0.04 | 0.03 | 0.42 | **0.38** | 0.28 | 0.23 | 0.38 |
| 2 | 0.01 | 0.13 | 0.33 | 0.47 | **0.07** | 0.25 | 0.07 | 0.07 |
| 3 | 0.07 | 0.27 | **0.46** | 0.03 | 0.17 | 0.38 | 0.32 | 0.46 |
| 4 | 0.06 | 0.35 | 0.02 | **0.37** | 0.19 | 0.35 | 0.32 | 0.37 |
| 5 | **0.39** | 0.27 | 0.17 | 0.10 | 0.06 | 0.44 | 0.38 | 0.39 |
| 6 | 0.21 | 0.05 | 0.06 | 0.14 | **0.54** | 0.33 | 0.29 | 0.54 |
| 7 | 0.03 | **0.51** | 0.02 | 0.21 | 0.22 | 0.41 | 0.35 | 0.51 |
| 8 | 0.14 | 0.04 | **0.74** | 0.07 | 0.00 | 0.35 | 0.25 | 0.74 |
| 9 | 0.11 | 0.05 | 0.08 | **0.72** | 0.04 | 0.44 | 0.35 | 0.72 |
| 10 | 0.03 | 0.00 | 0.55 | 0.08 | **0.34** | 0.25 | 0.17 | 0.34 |
| 11 | **0.04** | 0.10 | 0.17 | 0.22 | 0.47 | 0.00 | 0.01 | 0.04 |
| 12 | 0.41 | 0.19 | 0.10 | **0.20** | 0.10 | 0.41 | 0.21 | 0.20 |
| 13 | 0.02 | 0.06 | **0.82** | 0.02 | 0.08 | 0.33 | 0.20 | 0.82 |
| 14 | 0.18 | 0.22 | 0.13 | **0.41** | 0.07 | 0.52 | 0.43 | 0.41 |
| 15 | 0.02 | 0.12 | **0.49** | 0.32 | 0.04 | 0.31 | 0.22 | 0.49 |
| 16 | 0.06 | 0.18 | **0.57** | 0.15 | 0.04 | 0.17 | 0.14 | 0.57 |
| 17 | 0.08 | 0.09 | 0.23 | **0.43** | 0.17 | 0.41 | 0.32 | 0.43 |
| 18 | 0.00 | 0.02 | **0.46** | 0.50 | 0.01 | 0.29 | 0.18 | 0.46 |
| 19 | 0.03 | 0.13 | **0.62** | 0.10 | 0.12 | 0.38 | 0.29 | 0.62 |
| 20 | 0.17 | 0.10 | 0.06 | 0.51 | **0.14** | 0.10 | 0.03 | 0.14 |
| 21 | 0.03 | 0.03 | 0.25 | **0.52** | 0.16 | 0.27 | 0.19 | 0.52 |
| 22 | 0.03 | **0.44** | 0.09 | 0.02 | 0.42 | 0.33 | 0.27 | 0.44 |
| 23 | 0.12 | 0.07 | 0.09 | **0.40** | 0.32 | 0.36 | 0.26 | 0.40 |
| 24 | 0.47 | 0.08 | 0.13 | **0.24** | 0.06 | 0.43 | 0.29 | 0.24 |
| 25 | **0.05** | 0.60 | 0.27 | 0.06 | 0.01 | 0.20 | 0.05 | 0.05 |
| 26 | 0.44 | 0.07 | 0.06 | **0.40** | 0.04 | 0.42 | 0.32 | 0.40 |
| 27 | 0.05 | **0.73** | 0.07 | 0.02 | 0.13 | 0.39 | 0.30 | 0.73 |
| 28 | 0.45 | 0.03 | 0.16 | **0.24** | 0.10 | 0.13 | 0.06 | 0.24 |
| 29 | 0.39 | **0.19** | 0.11 | 0.17 | 0.10 | 0.22 | 0.16 | 0.19 |
| | | | | | Average | 0.32 | 0.23 | 0.41 |

Table V: Misconceptions found during interviews. Solid dots indicate misconceptions used most often. Hollow dots indicate misconceptions used less often.

| **Description** | | **Physical Aspects**<br><br>**Questions 10, 22,23** | **Current**<br><br>**Questions 8, 20** | **Energy**<br><br>**Question 3** | **Voltage**<br><br>**Questions 15, 16, 28, 29** |
|---|---|---|---|---|---|
| Battery superposition | 1 battery — bulb shines x bright 2 batteries, regardless of arrangement — bulb shines 2x bright | | | ○ | ● |
| Battery as a constant current source | Battery supplies same amount of current to each circuit regardless of the circuit's arrangement | ○ | | ● | ● |
| Complete circuit | Unable to identify a complete circuit — closed loop | | | | ○ |
| Contacts | Unable to identify the two contacts on the light bulb | ● | | | |
| Current consumed | Current value decreases as you move through circuit elements until you return to the battery where there is no more current left | ○ | ● | | ○ |
| Direct route | Battery is the only source of charge so only those elements with a direct contact to the battery will light | ○ | | | |
| E = 0 inside | Electric field inside a conductor is always zero | | ○ | | |
| I causes E | Current is the cause for the electric field inside the wires of the circuit | | ● | | |
| Local | Current splits evenly at every junction regardless of the resistance of each branch | ○ | | ○ | ● |
| $R_{eq}$ | Student equated the equivalent resistance of a circuit with an individual resistor | ○ | | | ○ |
| Resistive superposition | 1 resistor reduces the current by x 2 resistors reduce the current by 2x regardless of the resistor's arrangement | ○ | | | ○ |
| Rule application error | Misapplied a rule governing circuits. For example, used the equation for resistor in series when the circuit showed resistors in parallel | ○ | | | ○ |
| Sequential | Only changes before an element will affect that element | ○ | | | ○ |
| Term confusion I/R | Resistance viewed as being caused by the current. A resistor resists the current so a current must flow for there to be any resistance | ● | | | |
| Term confusion I/V | Voltage viewed as a property of current. Current is the cause of the voltage. Voltage and current always occur together | ❍ | | | ● |
| Topology | All resistors lined up in series are in series whether there is a junction or not. All resistors lined up geometrically in parallel are in parallel even if a battery is contained within a branch | ● | | | |
| $V=C_{eq}$ | Voltage calculated using equations for equivalent capacitance | | | ❍ | |
| $V=R_{eq}$ | Voltage calculated using equations for equivalent resistance | | | | ❍ |

Table VI: *t* -test results for each sample taking DIRECT version 1.0.

| Group | Mean and standard deviation for Males | Mean and standard deviation for Females | Degrees of freedom | *t* | *p* -value |
|---|---|---|---|---|---|
| Overall | 14 ± 4.7 | 12 ± 3.4 | 600 | 8.5 | $7.4 \times 10^{-17}$ |
| University | 16 ± 5.0 | 12 ± 3.7 | 123 | 5.2 | $4.6 \times 10^{-7}$ |
| High school | 13 ± 4.2 | 11 ± 3.3 | 425 | 5.7 | $1.1 \times 10^{-8}$ |

Figure 1: A circuit representing a series-parallel combination of equal resistances.

Figure 2: A circuit representing a parallel-series combination of equal resistances.

Figure 3: Distribution of scores for both versions of DIRECT - overall sample.

Figure 4: Question 7 from DIRECT version 1.1.

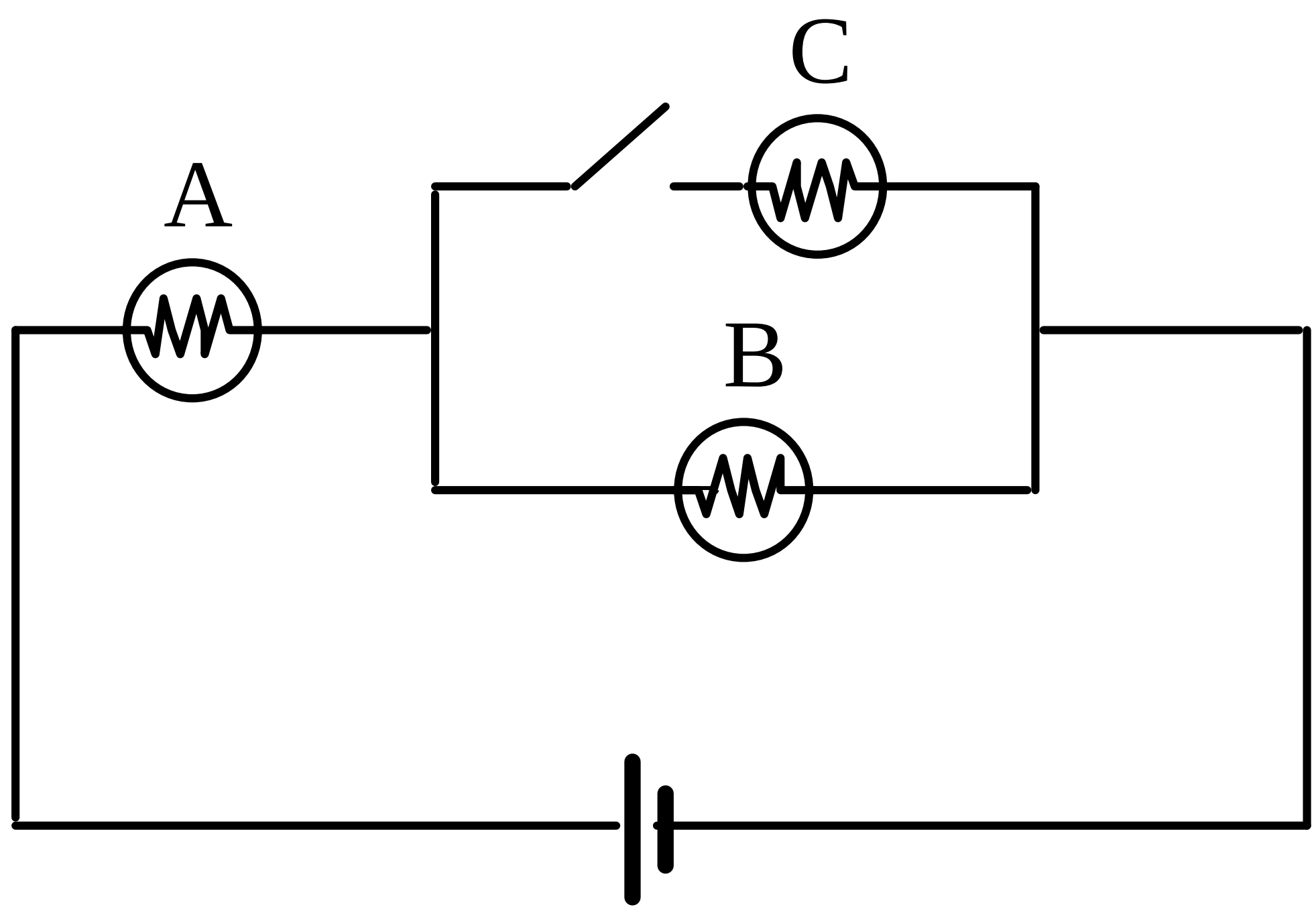

Fig. 1.

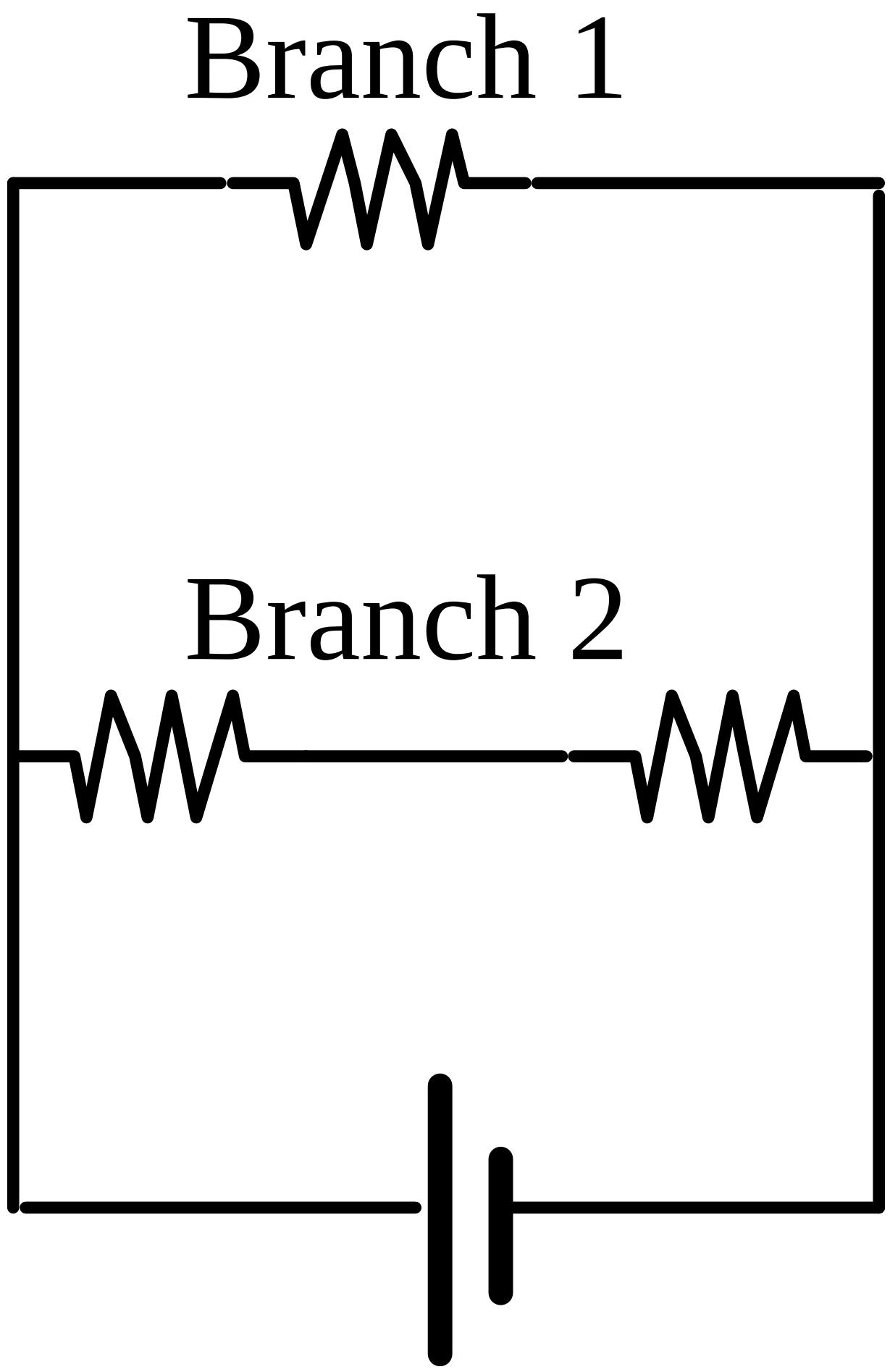

Fig. 2.

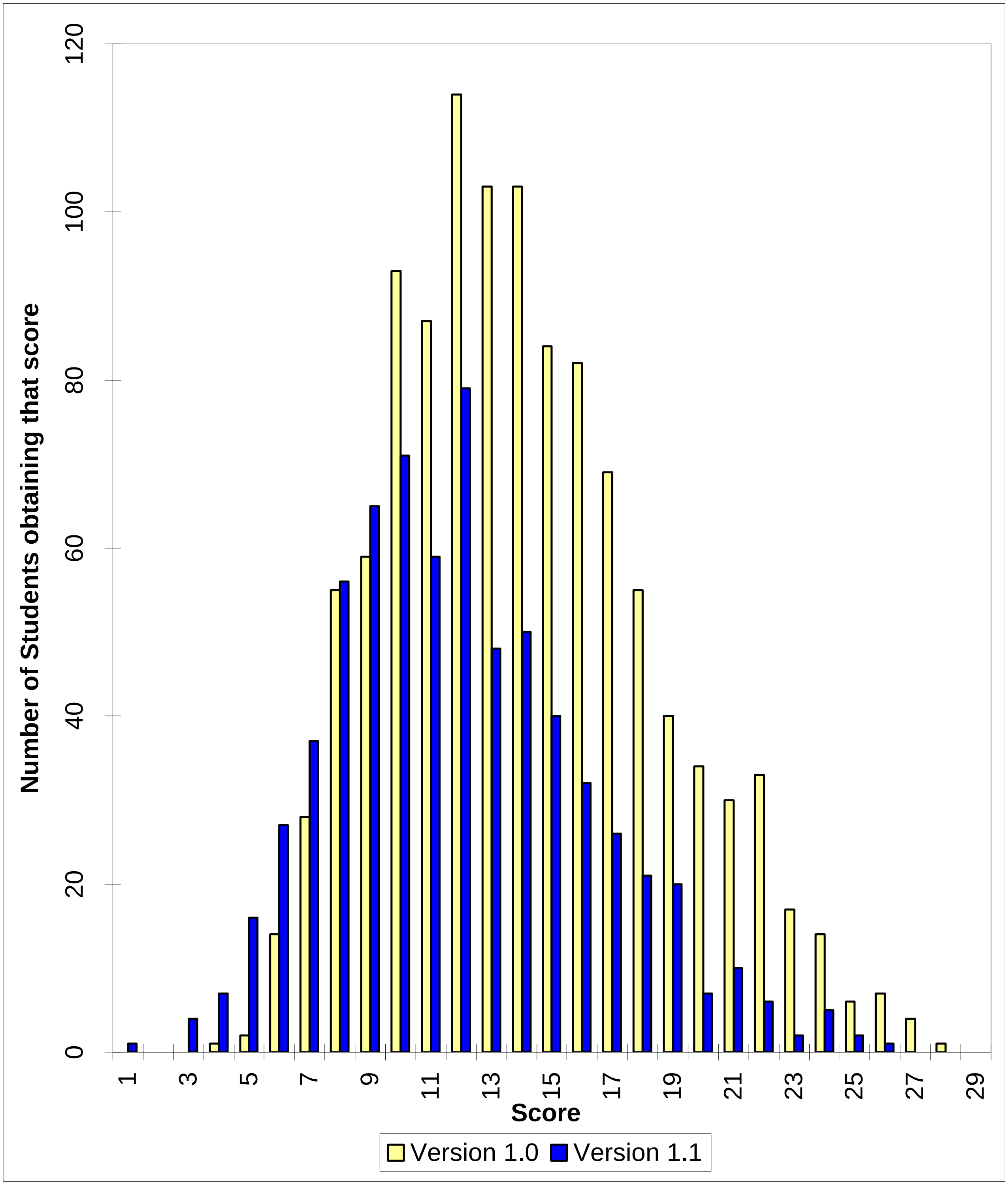

Fig. 3.

Compare the brightness of the bulb in circuit 1 with that in circuit 2. Which bulb is BRIGHTER?

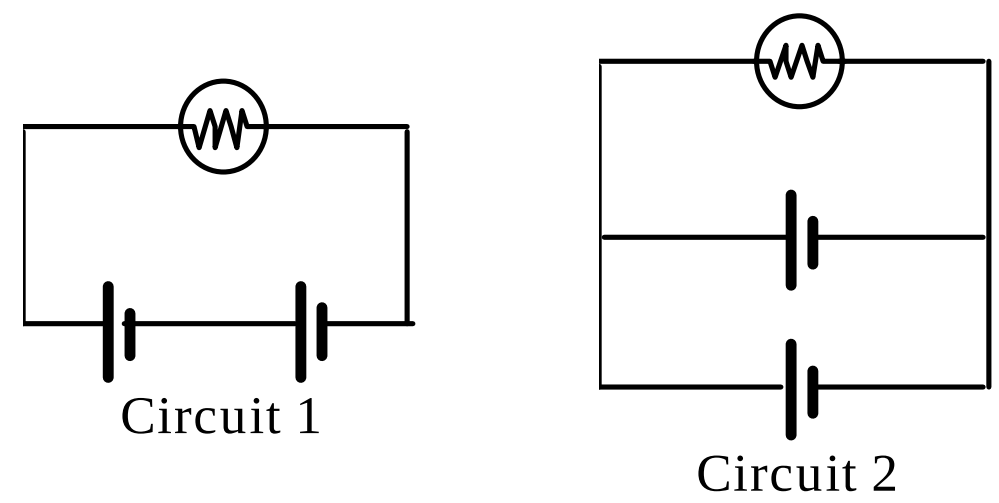

(A) Bulb in circuit 1 because two batteries in series provide less voltage

(B) Bulb in circuit 1 because two batteries in series provide more voltage

(C) Bulb in circuit 2 because two batteries in parallel provide less voltage

(D) Bulb in circuit 2 because two batteries in parallel provide more voltage

(E) Neither, they are the same

Fig. 4.